\documentclass[twocolumn]{IEEEtran}
\usepackage{epsfig}
\usepackage{epstopdf}
\usepackage{graphics}
\usepackage{lscape}			
\usepackage{listings}			
\usepackage{amsmath}	   
\usepackage{shadow}        
\usepackage{hhline}        
\usepackage{tabularx}      
\usepackage{latexsym}      
\usepackage{times}         
\usepackage{fancyhdr}      
\usepackage{graphicx}	   
\usepackage[font=small,labelfont=bf]{caption} 
\usepackage{float}		   
\usepackage{subcaption}	   
\usepackage[hyphens]{url}  
\usepackage{wrapfig}	   
\usepackage{cite}		   
\usepackage{capt-of}	   
\usepackage{verbatim}
\usepackage{multirow}  
\usepackage{xspace}		   
\usepackage[inline]{enumitem}
\usepackage[bookmarks=false]{hyperref}
\hypersetup{
  colorlinks = true,
  linkcolor = blue,
  anchorcolor = blue,
  citecolor = blue,
  filecolor = blue,
  urlcolor = blue
}

\usepackage{shortcuts}

\renewcommand{\paragraph}{\vspace{3pt}\noindent\textbf}
\begin{document}

\date{}

\setlength{\parskip}{3.5pt}

\title{The MalSource Dataset: Quantifying Complexity and Code Reuse in Malware Development}

\author{Alejandro Calleja, Juan Tapiador, and Juan Caballero%
\thanks{A. Calleja and J. Tapiador are with the
Department of Computer Science, Universidad Carlos III de Madrid, 28911 Leganes, Madrid, Spain.
E-mail: \{accortin, jestevez\}@inf.uc3m.es.}%
\thanks{J. Caballero is with the IMDEA Software Institute, Madrid, Spain.
E-mail: juan.caballero@imdea.org.}
}

\maketitle


\begin{abstract}
\noindent 
During the last decades, the problem of malicious and unwanted
software (malware) has surged in numbers and sophistication.
Malware plays a key role in most of today's cyber attacks and
has consolidated as a commodity in the underground economy.
In this work, we analyze the evolution of malware from 1975
to date from a software engineering perspective.
We analyze the source code of 456 samples from 428 unique families 
and
obtain measures of their size, code quality, and estimates of
the development costs (effort, time, and number of people).
Our results suggest an exponential increment of nearly one
order of magnitude per decade in aspects such as size and
estimated effort, with code quality metrics similar to those
of benign software.
We also study the extent to which code reuse is present in our dataset.
We detect a significant number of code clones across malware families and 
report which features and functionalities are more commonly shared.
Overall, our results support claims about the increasing complexity of malware 
and its production progressively becoming an industry.
\end{abstract}

\section{Introduction}
\label{sec:intro}

The malware industry seems to be in better shape than ever. In their 2015
Internet Security Threat Report~\cite{SymantecISTR20}, Symantec reports
that the total number of known malware in 2014 amounted to 1.7 billion,
with 317 million (26\%) new samples discovered just in the preceding year.
This translates into nearly 1 million new samples created every day. A
recent statement by Panda Security~\cite{Panda16} provides a proportionally
similar aggregate: out of the 304 million malware samples detected by their
engines throughout 2015, 84 million (27\%) were new. These impressive
figures can be partially explained by the adoption of reuse-oriented
development methodologies that make it exceedingly easy for malware
writers to produce new samples, and also by the increasing use of packers
with polymorphic capabilities. Another key reason is the fact that
over the last decade malware has become a profitable industry, thereby
acquiring the status of a \emph{commodity}~\cite{Caballero11,Grier12} in the
flourishing underground economy of cyber crime~\cite{Stringhini14,Thomas15}.
From a purely technical point of view, malware has experienced a
remarkable evolutionary process since the 1980s, moving from simple
file-infection viruses to stand-alone programs with network propagation
capabilities, support for distributed architectures based on rich
command and control protocols, and a variety of modules to execute
malicious actions in the victim. Malware writers have also rapidly
adapted to new platforms as soon as these acquired a substantial user base,
such as the recent case of smartphones~\cite{Suarez14}.

The surge in number, sophistication, and
repercussion of malware attacks has gone hand in hand with much
research, both industrial and academic, on defense and analysis techniques.
The majority of such investigations have focused on binary analysis,
since most malware samples distribute in this form. Only very
rarely researchers have access to the source code and can report
insights gained from its inspection. (Notable exceptions include the analysis
of the source code of 4 IRC bots by Barford and Yegneswaran~\cite{Barford2007inside}
and the work of Kotov and Massacci on 30 exploit kits~\cite{Kotov13anatomy}).
One consequence of the lack of wide availability of malware source code is
a poor understanding of the malware development process, its
properties as a software artifact, and how these
properties have changed in the last decades.

In this paper, we present a study of malware evolution from a
software engineering perspective. 
Our analysis is based on a dataset collected by the authors over two years
and composed of the source code of 456 malware samples ranging from
1975 to 2016.
Our dataset includes, among others, 
early viruses, worms, botnets, exploit kits, and remote access trojans (RATs).
This is the largest dataset of malware source code presented in
the literature.
We perform two separate analysis on this dataset. 
First, we provide quantitative measurements on the evolution of
malware over the last four decades.
Second, we study the prevalence of source code reuse among these 
malware samples.

To measure the evolution of malware complexity over time we use 
several metrics proposed in the software engineering community. 
Such metrics are grouped into three main categories:
(i) measures of size: number of
source lines of code (SLOC), number of source files, number of
different programming languages used, and number of function
points (FP); (ii) estimates of the development cost: 
effort (man-months), required time, and number of programmers;
and (iii) measures of code quality: comment-to-code ratio,
complexity of the control flow logic, and maintainability of the
code. 
We use these metrics to compare malware source code to a 
selection of benign programs. 

We also study the prevalence of source code reuse in our dataset. 
Code reuse--or code clone--detection is an important problem to 
detect plagiarism, copyright violations, and to preserve the 
cleanness and simplicity of big software projects~\cite{roy2009comparison}. 
Several authors have suggested that code cloning is a fairly common practice 
in large code bases, even if it also leads to bug propagation and 
poorly maintainable code~\cite{kamiya2002ccfinder}. 
Given the high amount of malware discovered on a daily basis, 
it is a common belief that most malware is not developed from scratch, 
but using previously written code that is slightly modified according to 
the attacker's needs~\cite{rahimian2014reverse}. 
Detecting clones in malware source code enables 
a better understanding of the mechanisms used by malware, 
their evolution over time, and 
may reveal relations among malware families.

This paper builds on our previous work that studied malware 
evolution using software metrics on a dataset of 151 malware samples 
covering 30 years~\cite{malsource}.
In this work, we present our updated dataset, 
which triples the number of original samples and extends the covered 
time frame to four decades.  
We redo the analysis on malware evolution to cover the new samples, and 
also provide a new analysis on malware source code reuse.

The main findings of our work include:
\begin{enumerate}
\item We observe an exponential increase of roughly one order of
magnitude per decade in the number of source code files and SLOC
and FP counts per sample. Malware samples from the 1980s and
1990s contain just one or a few source code files, are generally
programmed in one language and have SLOC counts of a few
thousands at most. Contrarily, samples from the late 2000s
and later often contain hundreds of source code files spanning
various languages, with an overall SLOC count of tens, and even
hundreds of thousands.
\item In terms of development costs, our estimates
evidence that malware writing has evolved from small projects of just
one developer working no more than 1-2 months full time, to
larger programming teams investing up to 6-8 months and, in some
cases, possibly more.
\item A comparison with selected benign
software projects reveals that the largest malware samples 
in our dataset present software metrics akin to those of
products such as \texttt{Snort} or \texttt{Bash}, but are
still quite far from larger software solutions.
\item The code quality metrics analyzed do not suggest
significant differences between malware and benign software.
Malware has slightly
higher values of code complexity and also better maintainability,
though the differences are not remarkable.
\item We find quite a large number of code reuse instances
in our dataset, specifically in C/C++ and Assembly code, that range
from a few lines to several thousands lines of code in length. 
An analysis of such clones reveals that commonly shared
functionality belongs to one of four groups:
\begin{enumerate}
\item Anti-analysis capabilities 
such as unpacking routines, polymorphic engines,
and code to kill antivirus (AV) processes.
\item Core malware artifacts, including shellcodes for initial infection, 
spreading routines, and code for various actions on the victim. 
\item Data clones such as arrays of passwords, process names, and IP addresses.
\item Data structures and associated functions, such as those 
needed to interact with PE or ELF files, 
popular communication protocols,
or the operating system kernel through documented and undocumented APIs.
\end{enumerate}
\end{enumerate}

The remaining of this paper is organized as follows.
In Section~\ref{sec:dataset} we describe our dataset of malware source code.
Section~\ref{sec:analysis} presents our quantitative measurements on the 
evolution of malware development. 
In Section~\ref{sec:code_sharing} we detail our code clone detection approach 
and results.
Section~\ref{sec:discussion} discusses the suitability of our approach, 
its limitations, and additional conclusions. 
Finally, Section~\ref{sec:conclusion} concludes the paper.

\section{Dataset}
\label{sec:dataset}

Our work is based on a dataset of malware source code collected by 
the authors over two years (2015--2016). 
Collecting malware source code is a challenging endeavor because  
malware is typically released in binary form.
Only occasionally its source code is released or leaked, 
with its availability being strongly biased towards 
classical viruses and early specimens.
When leaked, the source code may be difficult to access in underground forums. 
These challenges make it impossible to try to be complete. 
While we try to collect as many samples as possible, 
the goal is to acquire representative examples of the malware ecosystem 
during the last 40+ years, constrained to the limited availability. 

Samples were obtained from a variety of sources, 
including virus collection sites such as \textit{VX Heaven}~\cite{vxheaven}, 
code repositories such as \textit{GitHub}, 
classical e-zines published by historically prominent malware writing groups 
such as \textit{29A}, 
malware exchange forums, and 
through various P2P networks. 
We expanded our list of sources by using a snowballing methodology, 
exploring previously unknown sources that were referenced in sites 
under examination.

A sample in our dataset corresponds to a specific version of a 
malware project, 
where a malware project is most often referred to as a {\em malware family}.
A sample may comprise of one or multiple source code files typically bundled
as an archive (e.g., a ZIP file). 
Those files may be set in an arbitrarily complex directory structure 
and may be written in one or multiple programming languages 
(see Section~\ref{sec:analysis}).
Most often only one version of a family has been leaked, 
but occasionally we collect multiple, 
e.g., \texttt{Cairuh.A} and \texttt{Cairuh.B}.
For the vast majority of samples we do not know the author(s). 

\begin{figure}[t]
  \centering
  \includegraphics[width=\columnwidth]{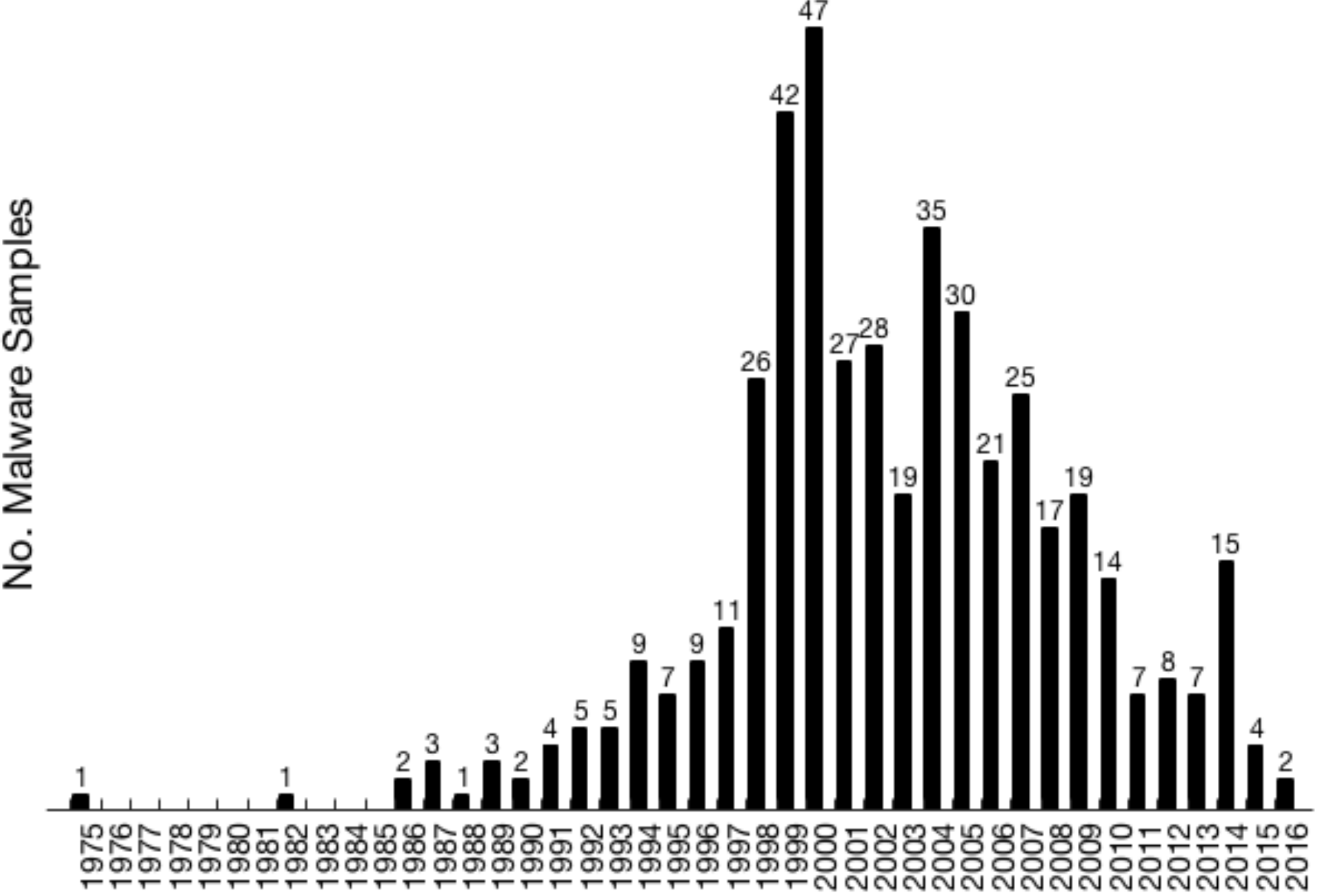}
  \caption{Distribution of malware source code samples in the dataset.}
  \label{fig:datasetHist}
\end{figure}

Our initial collection contained 516 samples. Each sample was first quickly verified through manual inspection and then compiled, executed and, whenever possible, functionally tested. At this point, 11.6\% of the obtained samples were discarded, either because testing them was unfeasible (e.g., due to nontrivial compilation errors or unavailability of a proper testing environment), or simply because they turned out to be fake. 

The 456 successfully tested samples that comprise our final dataset 
have been tagged with a year and a loose category. 
The year corresponds to their development when stated by the source, 
otherwise with the year they were first spotted in the wild. 
They are also tagged with a coarse-grained malware type:
Virus (V), Worm (W), Macro virus (M), Trojan (T), Botnet (B), RAT (R), 
Exploit kit (E), or Rootkit (K).
We are aware that this classification is rather imprecise. 
For instance, nearly all Botnets and RATs can be easily considered as Trojans, 
and, in some cases, show Worm features too. 
We chose not to use a more fine-grained malware type because it is 
not essential to our study and, furthermore, 
such classifications are problematic for many modern malware examples 
that feature multiple capabilities.

Figure~\ref{fig:datasetHist} shows the distribution by year of the final 
dataset of 456 samples. 
Approximately 61\% of the samples (281) correspond to the period 1995-2005. The second biggest set of samples (139) correspond to the period 2006-2016. Finally, the rest of samples (36) corresponds to the period ranging from 1975 to 1994.

The largest category is 
Virus (318 samples), 
followed by Worm (58), 
Botnet (26),
Trojan (26), 
RAT (12), 
Exploit kit (11),
Macro virus (4), and 
Rootkit (1).

\section{Malware Evolution Analysis}
\label{sec:analysis}

This section describes our analysis of the evolution of malware source code 
using software metrics.
It first quantifies the evolution in code size (Section~\ref{sec:codeSize}), 
then it estimates development cost (Section~\ref{sec:codeCost}), 
next it measures code quality (Section~\ref{sec:codeQuality}), and 
finally compares malware to benign code (Section~\ref{sec:comparison}).
In each section, we briefly introduce the software metrics used, 
and refer the reader to our original paper for more details~\cite{malsource}.

\begin{figure*}[t!]
    \centering
    \begin{subfigure}[b]{\columnwidth}
        \includegraphics[width=\textwidth]{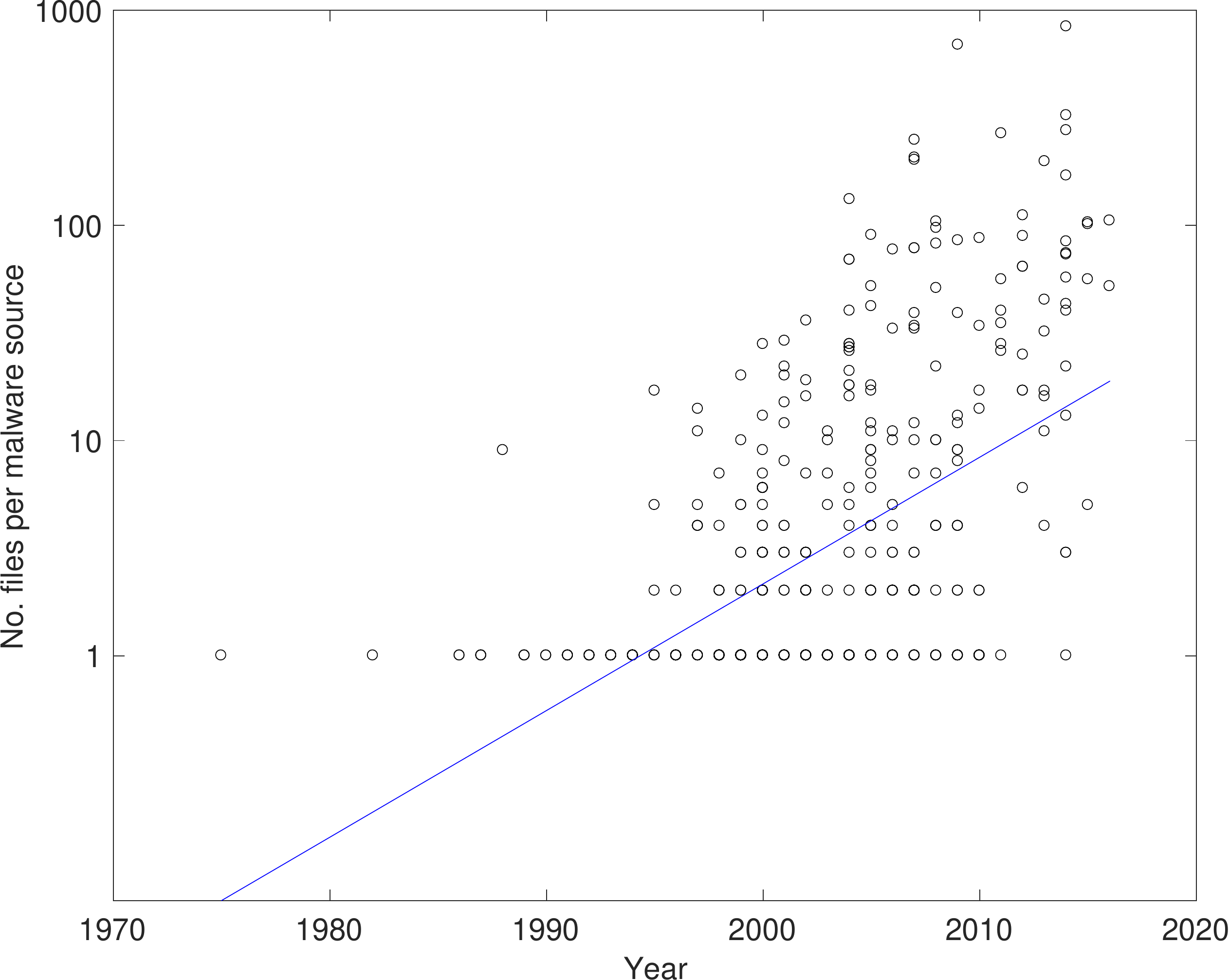}
        \caption{}
        \label{fig:numFiles}
    \end{subfigure}
~
    \begin{subfigure}[b]{\columnwidth}
        \includegraphics[width=\textwidth]{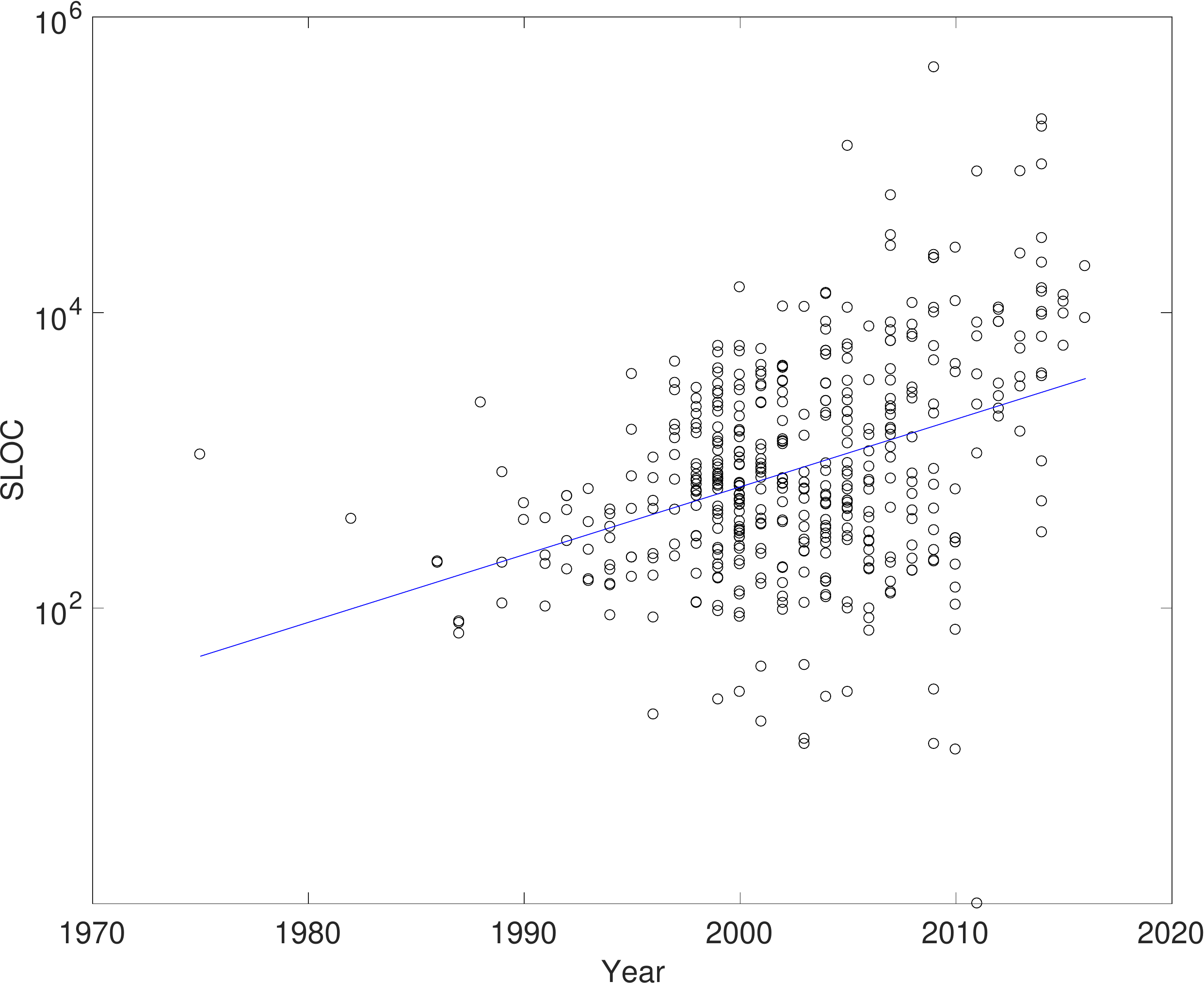}
        \caption{}
        \label{fig:numSloc}
    \end{subfigure}\hfill
    \caption{Number of files (a) and SLOC (b) for each sample in our
    dataset. Note that the y-axis is in logarithmic scale.}
\end{figure*}

\subsection{Code Size}
\label{sec:codeSize}

We use 3 different metrics to measure code size: 
number of files, number of source code lines, and 
function point estimates. 
We also measure the use of different programming languages in 
malware development.

\paragraph{Number of files.}
Figure~\ref{fig:numFiles} shows the distribution over time of the 
number of files comprising the source code of each sample in the dataset. 
Except for a few exceptions, until the mid 1990s there is
a prevalence of malicious code consisting of just one file. Nearly all
such samples are viruses written in assembly that, as discussed
later, rarely span more than 1,000 lines of code. 
This follows a relatively common practice of the 1980s and 1990s 
when writing short assembly programs.

From the late 1990s to date there is an exponential growth in
the number of files per malware sample. The code of viruses and
worms developed in the early 2000s is generally distributed across
a reduced ($<$10) number of files, while some Botnets and RATs from
2005 on comprise substantially more. 
For instance, Back Orifice
2000, GhostRAT, and Zeus, all from 2007, contain 206, 201, and 249
source code files, respectively. 
After 2010, no sample comprises a single file. 
Examples of this time period include KINS (2011),
SpyNet (2014), and the RIG exploit kit (2014) with 267, 324, and 838 files,
respectively. This increase reveals a more modular design,  
which also correlates with the use of higher-level programming languages 
discussed later, and the inclusion of more
complex malicious functionalities (e.g., network communications and
support for small databases).

Simple least squares linear regression over the data points shown
in Figure~\ref{fig:numFiles} yields a regression coefficient (slope)
of 1.14. (Note that the y-axis is in logarithmic scale and, therefore,
such linear regression actually corresponds to an exponential fit.)
This means that the number of files has grown at an approximate
yearly ratio of 14\%, i.e., it has doubled every 5 years.

\paragraph{Number of lines.}
Traditionally, the number of lines in the source code of a program, 
excluding comment and blank lines (SLOCs),
has been employed as the most common metric for measuring its size. 
In our analysis we use \texttt{cloc}~\cite{cloc}, 
an open-source tool that counts SLOCs, blank lines, and comment lines, 
and reports them broken down by programming language. 
Figure~\ref{fig:numSloc} shows the SLOCs for each sample, 
obtained by simply aggregating the SLOCs of all source 
code files of the sample, irrespective of 
the programming language in which they were written.

Again, the growth over the last 40 years is clearly exponential.
Up to the mid 1990s viruses and early worms rarely exceeded
1,000 SLOCs. Between 1997 and 2005 most samples contain several
thousands SLOCs, with a few exceptions above that figure, e.g.,
Simile (10,917 SLOCs) or Troodon (14,729 SLOCs). The increase in SLOCs
during this period correlates positively with the number of
source code files and the number of different programming languages
used. Finally, a significant number of samples from 2007 on exhibit
SLOCs in the range of tens of thousands. For instance,
GhostRAT (33,170), Zeus (61,752), KINS (89,460), Pony2 (89,758), or
SpyNet (179,682). Most of such samples correspond to moderately
complex malware comprising of more than just one executable.
Typical examples include Botnets or RATs featuring a web-based C\&C
server, support libraries, and various types of bots/trojans. There
are exceptions, too. For instance, Point-of-Sale (POS) trojans such as
Dexter (2012) and Alina (2013) show relatively low SLOCs 
(2,701 and 3,143, respectively).

In this case the linear regression coefficient over the data points
is 1.11, i.e., the SLOCs per malware have increased
approximately 11\% per year; or, equivalently, the SLOC count doubles
every 6.5 years, resulting in an increase of nearly an order of
magnitude each decade.

\begin{figure}[t]
  \includegraphics[width=\columnwidth]{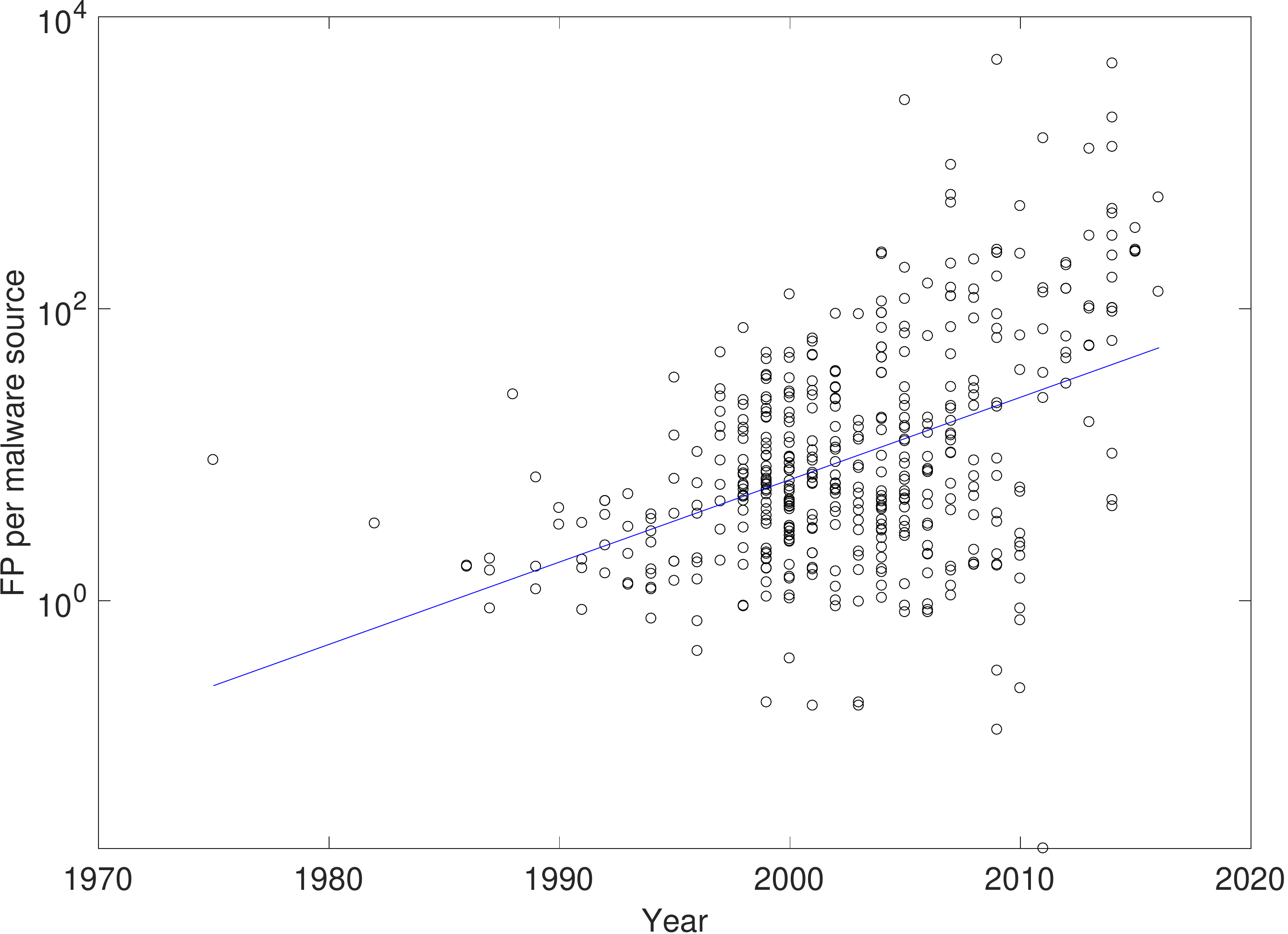}
  \caption{Function point counts for each sample in the dataset. 
  Note that the y-axis is in logarithmic scale.}
  \label{fig:numFP}
\end{figure}

\paragraph{Function points estimates.}
Although SLOCs is the most popular metric for measuring 
project size, it has a number of shortcomings~\cite{Nguyen07}.
Most notably, when comparing 
the sizes of projects developed using different programming 
languages, SLOCs may lead to misleading conclusions since this metric does not
take into account the  programming language expressiveness. 
To address this issue, we leverage the 
\emph{function-point count}~\cite{Albrecht79,Albrecht83} (FPC) metric,
which aims to capture the overall functionality of the software.
The function-point count is measured using four program features: 
external inputs and outputs, user interactions, 
external interfaces, and files used. 
The expected size in SLOCs of a software project can be estimated 
(before it is coded) from 
function-point counts through a process known as 
\emph{backfiring}~\cite{Jones95}. 
This process uses programming languages empirical tables (PLTs)
that provide the average SLOCs per function point
for different programming languages. 
In our analysis, we use a reversed backfiring process that uses  
PLT v8.2~\cite{Jones96} to obtain function-point counts from SLOCs. 
We use those function-point counts as a normalized size 
for malware written in different languages. 

Figure~\ref{fig:numFP} shows, as expected, 
a clear correlation between FPC and SLOCs and  
the conclusions in terms of sustained growth are similar. 
Starting in 1990, there is roughly an increase of one
order of magnitude per decade. Thus, in the 1990s most early
viruses and worms contain just a few ($<10$) function points. From 2000
to 2010 the FPC concentrate between 10 and 100,
with Trojans, Botnets, and RATs accounting for the higher
counts. Since 2007, many samples exhibit FPC 
of 1,000 and higher; examples include Pony2 (2013), with
1,240, SpyNet (2014), with 2,028, and the RIG exploit kit
(2014), with 4,762. Linear regression over the data
points yields a coefficient of 1.13, i.e., FPCs per malware
have suffered an approximate growth of 13\% per year; or,
equivalently, FPCs double every 5.5 years.

\begin{figure*}[t]
    \centering
    \begin{subfigure}[b]{.50\textwidth}
    \centering
        \includegraphics[width=\textwidth]{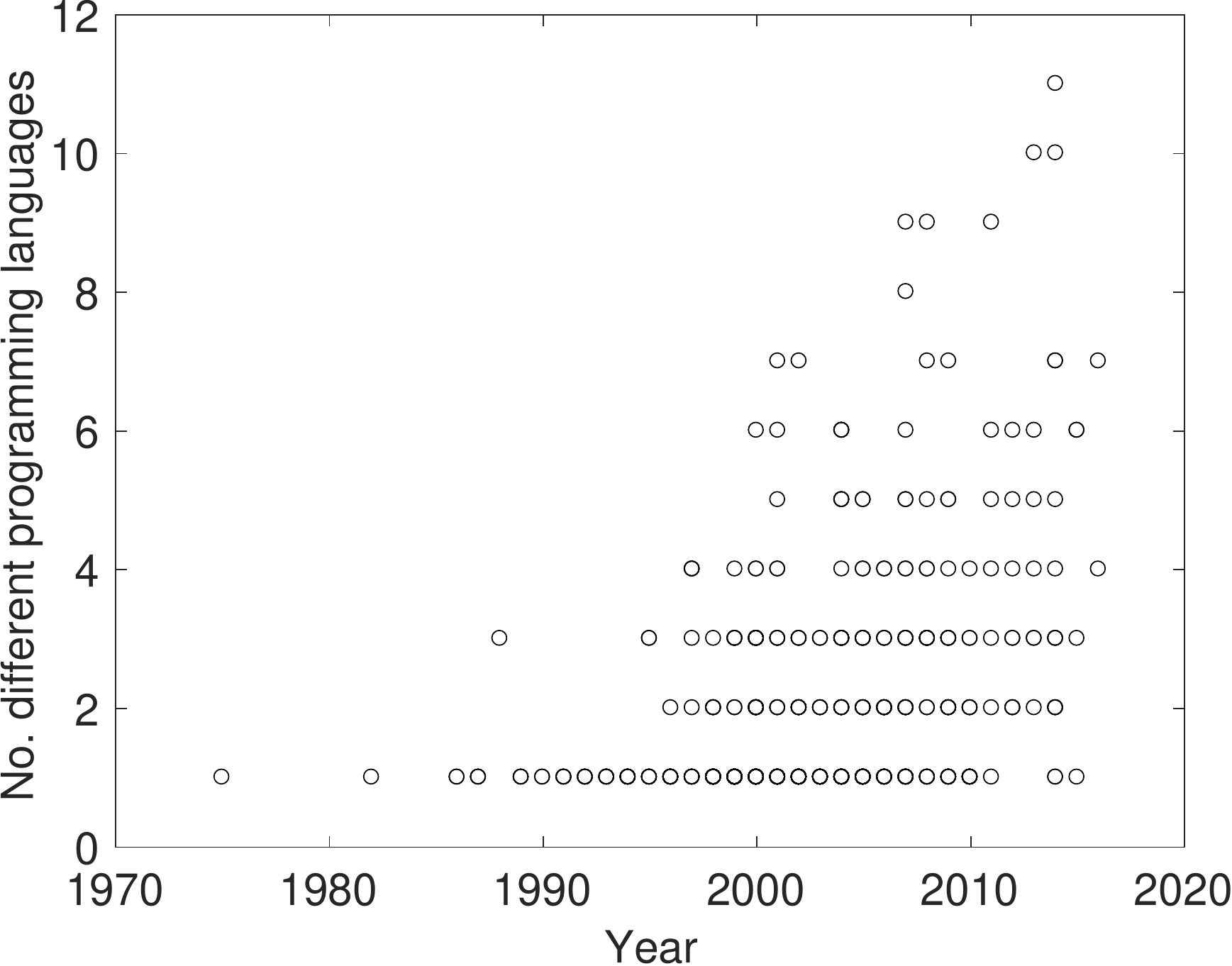}
        \caption{}
        \label{fig:numLang}
    \end{subfigure}%
\hfill%
    \begin{subfigure}[b]{.50\textwidth}
    \centering
        \includegraphics[width=\textwidth]{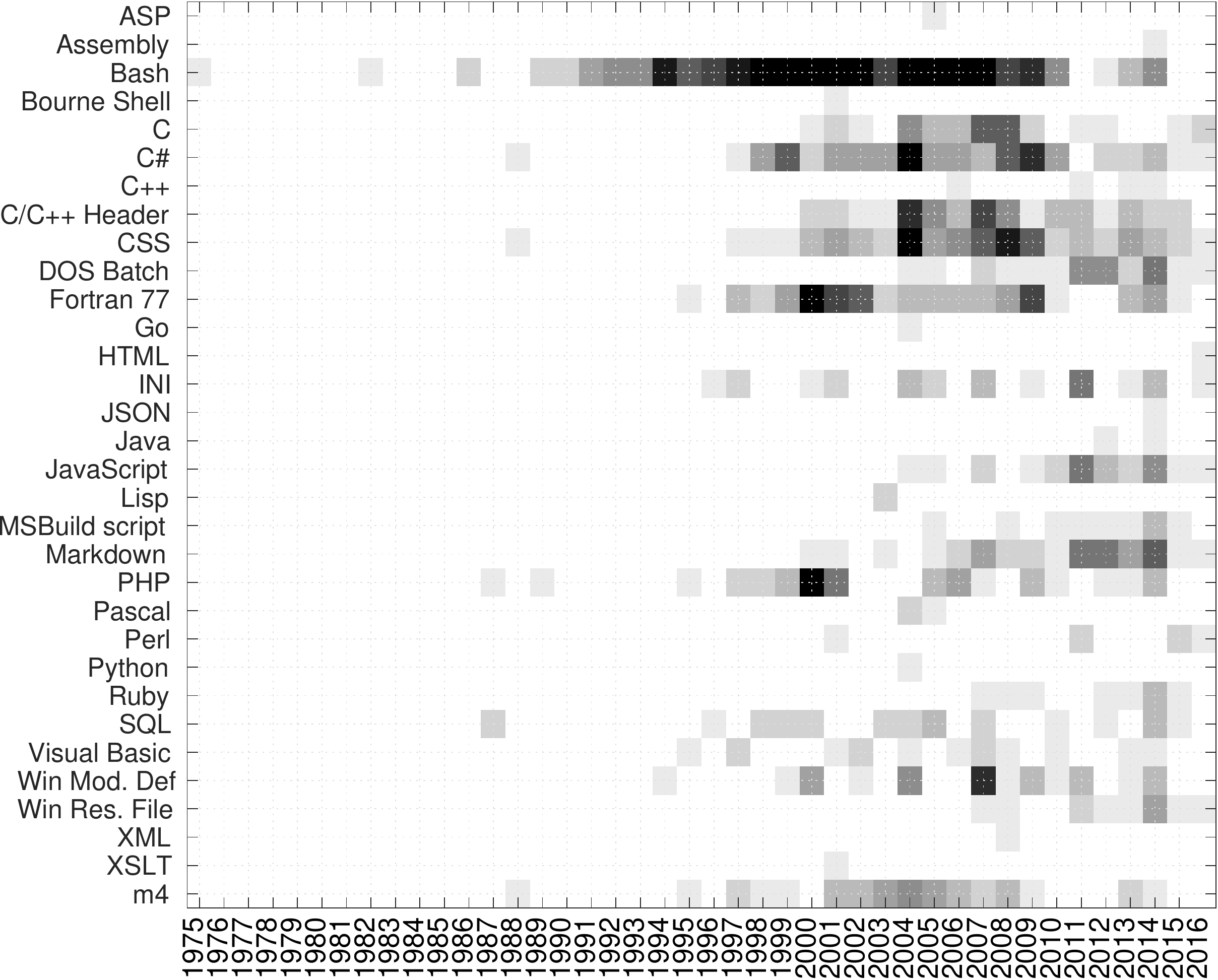}
        \caption{}
        \label{fig:langUsage}
    \end{subfigure}%
    \caption{\textbf{(a)} Number of programming languages
    per malware sample in the dataset. Darker circles represent overlapping 
    data points. \textbf{(b)} Use of programming
    languages in malware samples. The chart shows the number of samples
    using a particular language each year, with darker colors representing
    higher number of samples.
    }
    \label{fig:langs}
\end{figure*}

\paragraph{Programming languages.}
Figure~\ref{fig:numLang} shows the distribution over time of the 
number of different languages used to develop each malware sample.
This includes not only compiled and interpreted languages such as
assembly, C/C++, Java, Pascal, PHP, Python, or Javascript, but also
others used to construct resources that are part of the final
software project (e.g., HTML, XML, CSS) and scripts used to build
it (e.g., BAT or Make files).

Figure~\ref{fig:langUsage} shows the usage of different programming
languages to code malware over time in our dataset.
The pattern reveals the prevalent use of assembly until the late
2000s. From 2000 on, C/C++ become increasingly popular, as well as
other ``visual'' development environments such as Visual Basic
and Delphi (Pascal). Botnets and RATs from 2005 on also make
extensive use of web interfaces and include numerous HTML/CSS
elements, pieces of Javascript, and also server-side functionality
developed in PHP or Python. Since 2012 the distribution
of languages is approximately uniform, revealing the heterogeneity
of technologies used to develop modern malware.

\subsection{Development Cost}
\label{sec:codeCost}

An important problem in software engineering is to make an accurate
estimate of the cost required to develop a 
software system~\cite{Sommerville06}. 
A prominent approach to this problem are algorithmic cost modeling methods, 
which provide cost figures using as input various project properties such as 
code size and organizational practices. 
Probably the best known of these methods is the Constructive Cost Model 
(COCOMO)~\cite{COCOMO}. COCOMO is an empirical model
derived from analyzing data collected from a large number of software
projects. 
COCOMO provides the following equations for estimating three metrics related 
to the cost of a project: 
effort (in man-months), 
development time (in months), 
and number of people required. 

\begin{equation}\label{Eq:COCOMO_Effort}
E = a_b (\mathrm{KLOC})^{b_b}
\end{equation}

\begin{equation}\label{Eq:COCOMO_DevTime}
D = c_b E^{d_b}
\end{equation}

\begin{equation}\label{Eq:COCOMO_People}
P = \frac{E}{D}
\end{equation}

In the equations above, KLOC represent the estimated SLOCs 
in thousands and $a_b$, $b_b$, $c_b$, $d_b$ are empirically obtained 
regression coefficients provided by the model. 
The value of these coefficients depends on the nature of the project. 
COCOMO considers three different types of projects: (i) \emph{Organic} projects (small programming team, good
experience, and flexible software requirements); \emph{Semi-detached}
projects (medium-sized teams, mixed experience, and a combination of
rigid and flexible requirements); and (iii) \emph{Embedded} projects
(organic or semi-detached projects developed with tight constraints). 
For our analysis, we decided to consider all samples as organic
for two reasons. First, it is reasonable to assume that,
with the exception of a few cases, malware development
has been led so far by small teams of experienced programmers. 
Additionally, we favor a conservative estimate of development cost, 
which is achieved using the lowest COCOMO coefficients 
(i.e., those of organic projects). 
Thus, our estimates can be seen as a 
(estimated) lower bound of development cost.

\begin{figure*}[t!]
    \centering
    \begin{subfigure}[b]{.5\textwidth}
    \centering
        \includegraphics[width=\textwidth]{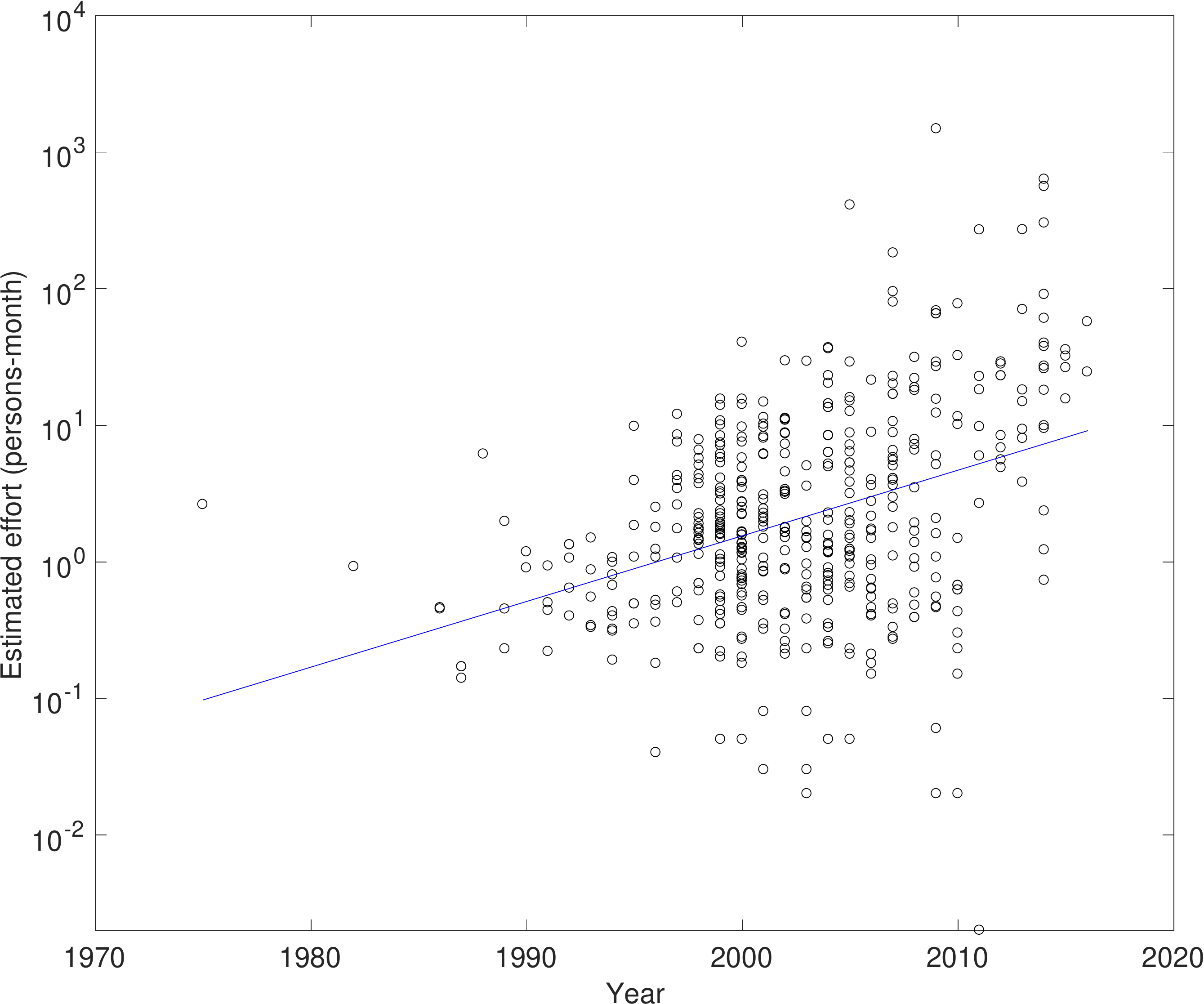}
        \caption{}
        \label{fig:devEffort}
    \end{subfigure}%
\hfill
    \begin{subfigure}[b]{.5\textwidth}
    \centering
        \includegraphics[width=\textwidth]{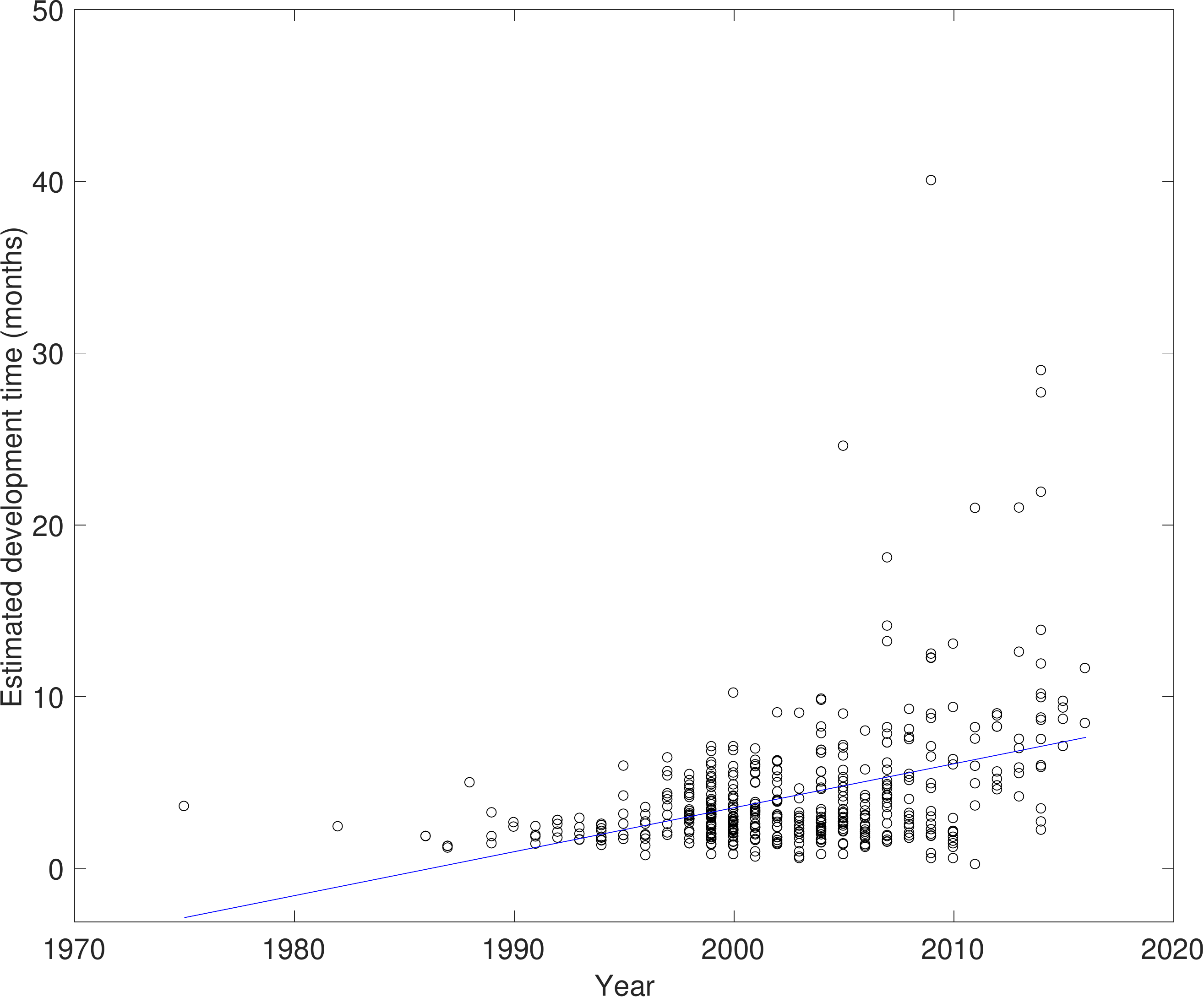}
        \caption{}
        \label{fig:devTime}
    \end{subfigure}%
    
    \begin{subfigure}[b]{.5\textwidth}
     \centering
        \includegraphics[width=\textwidth]{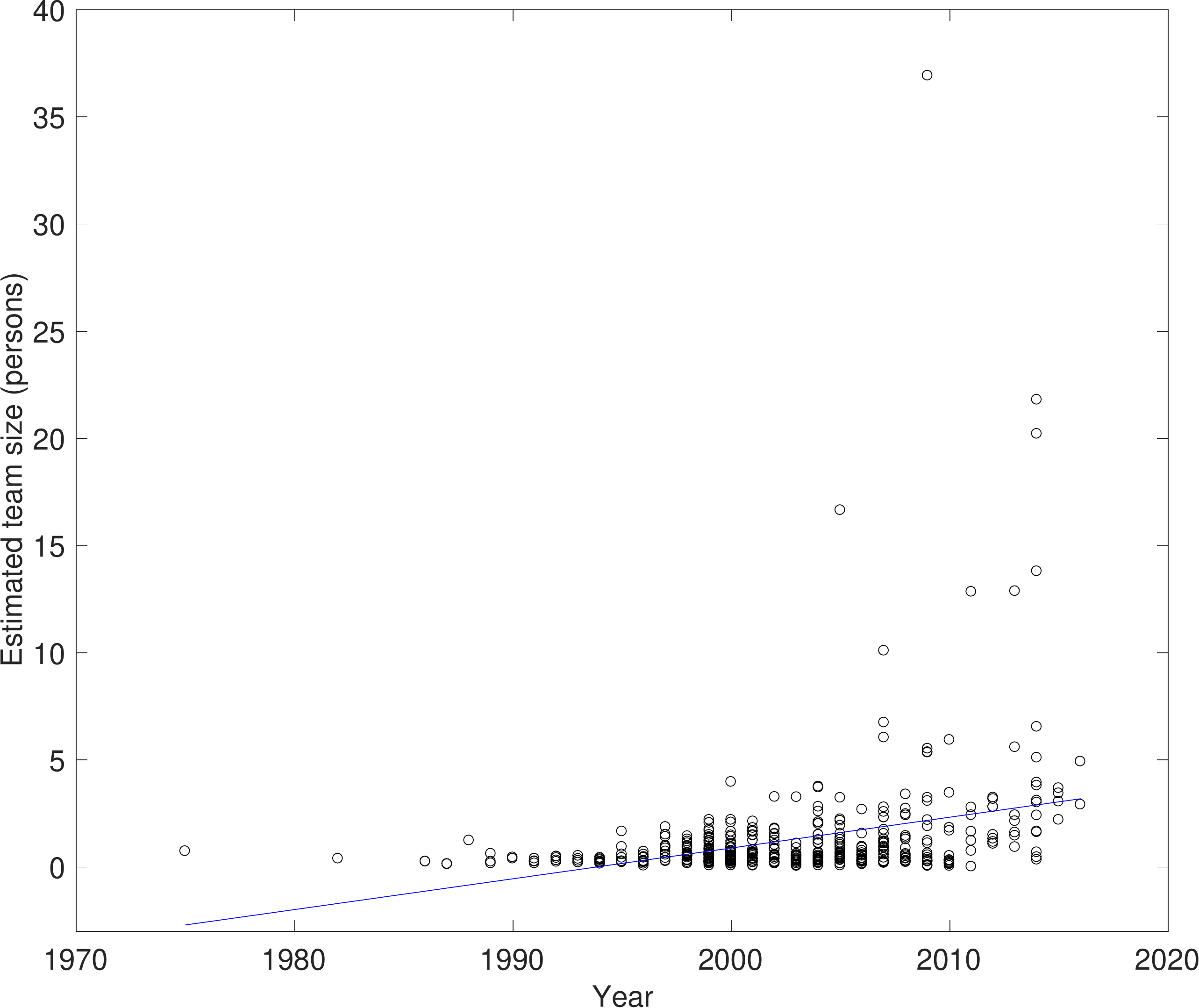}
        \caption{}
		\label{fig:teamSize}
    \end{subfigure}
\hfill
    \begin{subfigure}[b]{.48\textwidth}
    \centering
    	\begin{tabular}{|l|c|r|r|r|}
    	\hline
    	\multicolumn{1}{|c|}{\textbf{Sample}} &
    	\textbf{Year} &
    	\multicolumn{1}{c|}{\textbf{E}} &
    	\multicolumn{1}{c|}{\textbf{D}} &
    	\multicolumn{1}{c|}{\textbf{P}}\\
    	\hline\hline
		Anthrax & 1990 & 0.9 & 2.40 & 0.38\\
		Batvir & 1994 & 0.40 & 1.76 & 0.23\\
		AIDS & 1999 & 0.23 & 1.43 & 0.16\\
		IISWorm & 1999 & 0.41 & 1.78 & 0.23\\
		ILOVEYOU & 2000 & 0.44 & 1.83 & 0.24\\
		Blaster & 2003 & 1.48 & 2.90 & 0.51\\
		Sasser & 2004 & 2.27 & 3.41 & 0.67\\
		Mydoom & 2004 & 8.35 & 5.60 & 1.49\\
		GhostRAT & 2007 & 94.84 & 14.10 & 6.73\\
    Zeus & 2007 & 182.14 & 18.07 & 10.08\\
		KINS & 2011 & 358.39 & 23.37 & 15.34\\
		Dexter & 2012 & 9.08 & 5.78 & 1.57\\
		Dendroid & 2014 & 37.65 & 9.93 & 3.79\\
    Tinba & 2014 & 39.84 & 10.14 & 3.93\\
		Mirai & 2016 & 24.48 & 8.43 & 2.9\\
		Mazar & 2016 & 57.16 & 11.63 & 4.91\\
    	\hline
    	\multicolumn{1}{c}{~}\\
    	\end{tabular}
      \caption{}
      \label{fig:COCOMOexamples}
    \end{subfigure}%
    \caption{COCOMO cost estimators for the malware samples in the dataset.
    \textbf{(a)}  Effort (man-months). \textbf{(b)} Development time (months).
    \textbf{(c)} Team size (number of people). \textbf{(d)} Selected examples
    with effort (E), development time (D), and number of people (P). Note that
    in \textbf{(a)} and \textbf{(b)} the y-axis is shown in logarithmic scale.}
    \label{fig:CCMestimators}
\end{figure*}

Figure~\ref{fig:CCMestimators} shows the COCOMO estimates for the effort,
time, and team size required to develop the malware samples in
our dataset. Figure~\ref{fig:devEffort} shows the COCOMO estimation of effort 
in man-months. The evolution over time
is clearly exponential, with values roughly growing one order of
magnitude each decade. While in the 1990s most samples required
approximately one man-month, this value rapidly escalates up
to 10--20 man-months in the mid 2000s, and to 100s for a few samples in the
last years. Linear regression confirms this, yielding a regression
coefficient of 1.11; i.e., the effort growth ratio per year is
approximately 11\%; or, equivalently, it doubles every 6.5 years.

The estimated time to develop the malware samples
(Figure~\ref{fig:devTime}) experiences a linear increase up to
2010, rising from 2-3 months in the 1990s to 7-10 months
in the late 2000s. The linear regression coefficient in this case
is 0.255, which translates into an additional month every 4 years.
Note that a few samples from the last 10 years report a
considerable higher number of months, such as Zeus (2007) or
SpyNet (2014) with 18.1 and 27.7 months, respectively.

The amount of people required to develop each sample
(Figure~\ref{fig:teamSize}) grows similarly. Most early
viruses and worms require less than one person (full time). From
2000 on, the figure increases to 3-4 persons for some samples.
Since 2010, a few samples report substantially higher estimates. 
For these data, the linear regression coefficient is 0.143,
which roughly translates into an additional team member every
7 years.

Finally, the table in Figure~\ref{fig:COCOMOexamples} provides
some numerical examples for a selected subset of samples.

\subsection{Code Quality}
\label{sec:codeQuality}

We measure 2 aspects of code quality: 
source code complexity and 
software maintainability.

\begin{figure}[h]
  \includegraphics[width=\columnwidth]{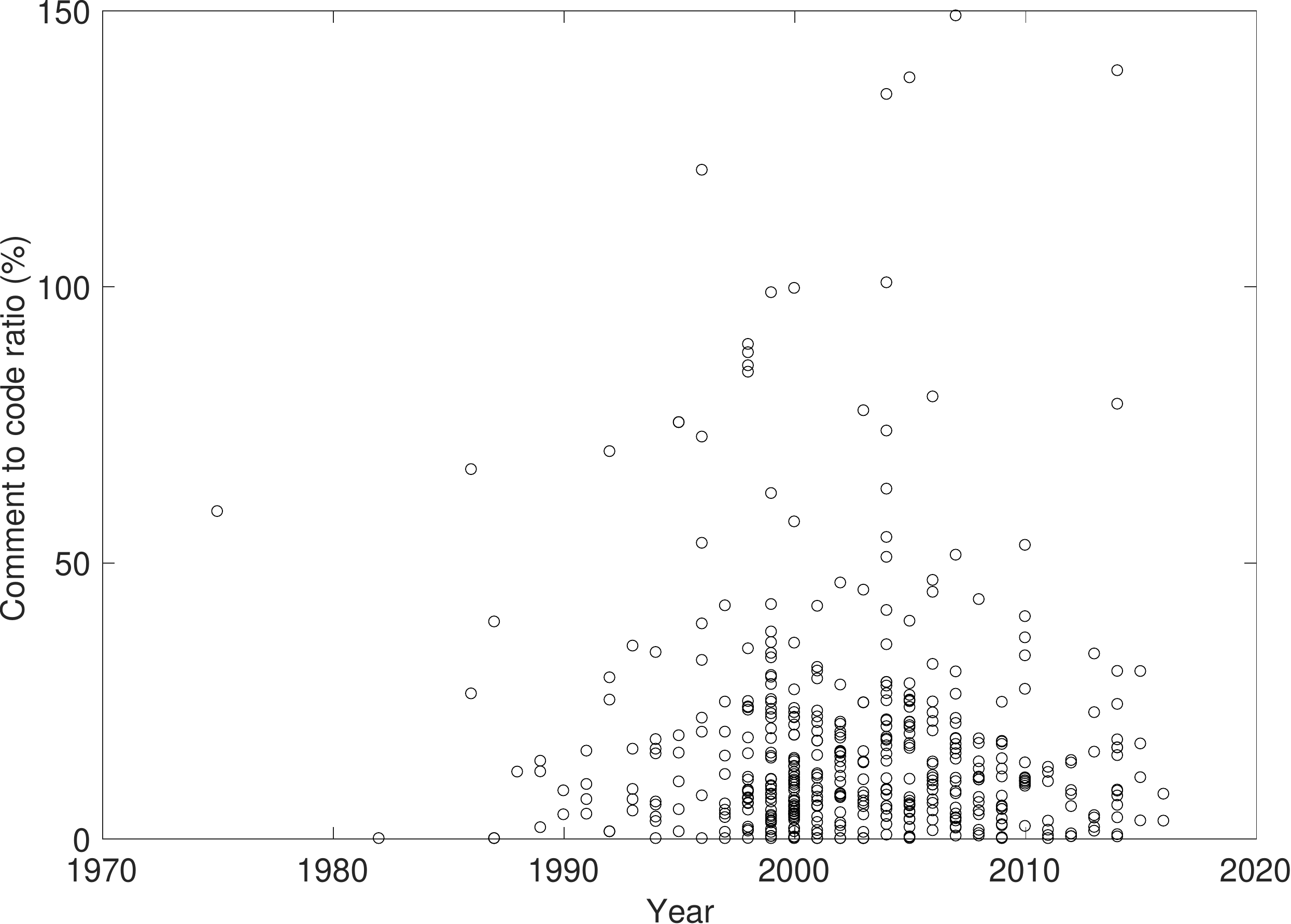}
  \caption{Comments-to-code ratio for each sample in the dataset.}
  \label{fig:commentsRatio}
\end{figure}

\paragraph{Complexity.}
To measure software complexity 
we use McCabe's cyclomatic complexity~\cite{McCabe76},
one of the earliest---and still most widely used---software complexity metrics. 

Despite having been introduced more than 40 years ago, it is still regarded as
a useful metric to predict how defect-prone a software piece is
\cite{ebert2016cyclomatic}, hence its use in many software measurement studies. 
For instance Warren et al. \cite{warren1999evolution} characterized the
evolution of modern websites by measuring different parameters, including the
complexity of their source code. More recently, Hecht et al. included McCabe's
complexity in their analysis of the complexity evolution of Android
applications \cite{hecht2015tracking}.

The cyclomatic complexity (CC) of a piece of source code is computed 
from its control flow graph (CFG) and measures the number of linearly 
independent paths within the CFG.
Mathematically, the cyclomatic complexity is given by:
\begin{equation}\label{Eq:CyclomaticComplexity}
CC = E - N + 2P
\end{equation}

\noindent where $E$ is the number of edges in the CFG, $N$ the number of nodes, 
and $P$ the number of connected components. 
There are many available tools for measuring this 
metric~\cite{JHAWK,Radon,EclipseMetrics}, but most of them 
only support a small subset of programming languages. 
To compute the cyclomatic complexity we use 
the Universal Code Count (\texttt{UCC})~\cite{UCC}.
\texttt{UCC} works over C/C++, C\#, Java, SQL, Ada, Perl, ASP.NET, JSP, CSS,
HTML, XML, JavaScript, VB, PHP, VBScript, Bash, C Shell Script,
Fortran, Pascal, Ruby, and Python. Since our dataset contains source
code written in different languages, \texttt{UCC} best suits
our analysis. Still, it limited our experiments since it is not compatible with assembly source code, 
which appears in many projects in our dataset (see Figure~\ref{fig:langUsage}). 
Filtering out samples that contain at least one source file in assembly 
left 144 samples for this analysis, i.e., 32\% of the dataset.

\begin{figure}[t]
    \centering
    \begin{subfigure}[b]{.50\textwidth}
    \centering
        \includegraphics[width=\columnwidth]{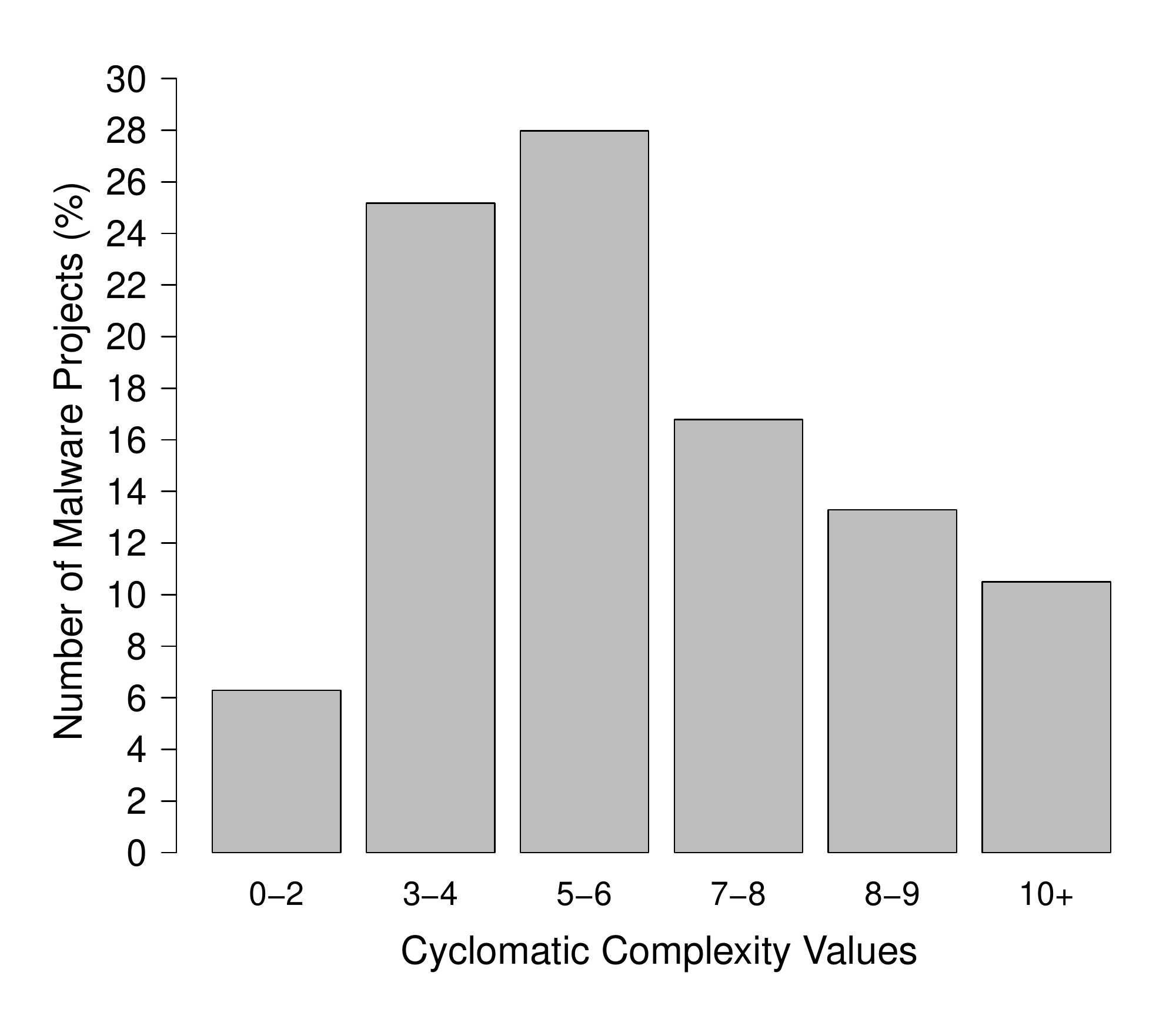}
        \caption{}
        \label{fig:complexityMalware}
    \end{subfigure}%
\hfill%
    \begin{subfigure}[b]{.50\textwidth}
    \centering
        \includegraphics[width=\columnwidth]{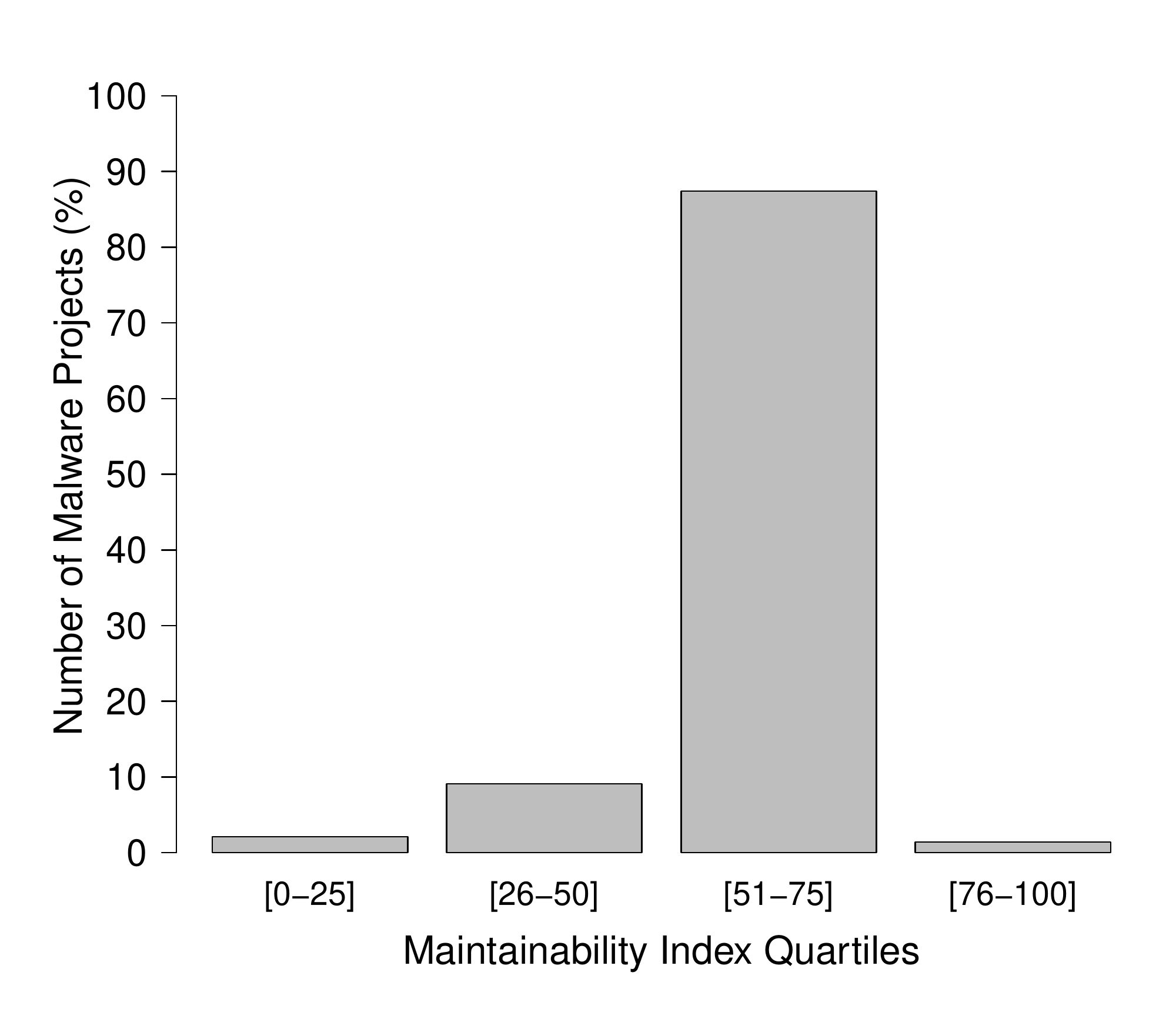}
        \caption{}
        \label{fig:maintainabilityMalware}
    \end{subfigure}

    \caption{Distributions of cyclomatic complexity \textbf{(a)} and maintainability index \textbf{(b)} for malware samples in the dataset.}
\end{figure}

Figure~\ref{fig:complexityMalware} shows the distribution of
the average cyclomatic complexity per function for each analyzed sample.
Most of the samples have functions with complexities between 3 and 8,
with values in the interval $[5,6]$ being the most common. Overall,
this can be seen as a supporting evidence of a generally modular
design with a good break down into fairly simple functions and class
methods. There are, however, some counterexamples. We observed a
number of functions with complexity higher than 10, which exceeds
McCabe's recommended complexity threshold.

\paragraph{Maintainability.}
A concept often linked to software quality is source code maintainability.
Maintainability is connected to complexity, since high complexity
translates into poor maintainability~\cite{Lehman96}. 
Maintaining a software product generally involves fixing bugs and 
adding new features. The documentation found in the source code as 
code comments can have a great impact in facilitating this process. Thus,
the comments-to-code ratio (or simply ``comments ratio'') has traditionally
been the metric used to measure documentation quality
\cite{Oman92,garcia1996maintainability}.

Figure~\ref{fig:commentsRatio}
shows the comments-to-code ratio for each sample, 
computed as the number of comment lines
divided by the SLOCs. There is no clear pattern in the data, which
exhibits an average of 17.2\%, a standard deviation of 21.5\%,
and a median value of 10.7\%. There are a few notable outliers,
though. For example, W2KInstaller (2000) and OmegaRAT (2014) 
show ratios of 99.6\% and 139.1\%, respectively. Conversely,
some samples have an unusually low comments ratio. We ignore
if they were originally developed in this way or, perhaps, comments
were cut off before leaking/releasing the code.

A more direct metric for measuring the maintainability of a software project is 
the maintainability index ($MI$)~\cite{Oman92}. 
This metric 
is a quantitative estimator of how easy is to understand, support, and modify 
the source code of a project. 
A popular definition of $MI$ is:
\begin{equation}\label{Eq:MaintainabilityIndex}
MI  = 100 \frac{171 - 5.2\ln{(\overline{V})} - 0.23 \overline{M} - 16.2 \ln{(\overline{SLOC})}}{171}
\end{equation}

\noindent where $\overline{V}$ is Halstead’s average volume per module (another classic complexity metric; see~\cite{Halstead77} for details), $\overline{M}$ is the average cyclomatic complexity per module, and $\overline{SLOC}$ is the average number of source code lines per module. 
$MI$ has been included in Visual Studio since 2007 \cite{metrics_visual_studio}.
Visual Studio flags modules with $MI<20$ as difficult to maintain. 

We use Equation~(\ref{Eq:MaintainabilityIndex}) for 
computing an $MI$ upper bound for each sample in our dataset. 
Note that we cannot estimate $MI$ exactly since 
we do not have the average Halstead's volume for each sample. 
Since this is a negative factor in Equation~(\ref{Eq:MaintainabilityIndex}), 
the actual $MI$ value would be lower than our estimate. Nevertheless, note
that such factor contributes the lowest, so we expect
our estimate to provide a fair comparison among samples.

Figure~\ref{fig:maintainabilityMalware} shows the distribution
of $MI$ values grouped in quartiles. 
Most samples have an $MI$ in the third quartile, 
and only 15 samples fall short of the recommended threshold of $MI<20$, 
mainly because of a higher-than-expected cyclomatic complexity.

\begin{table*}[t]
\centering
\begin{tabular}{|l|l|c|r|r|r|r|r|r|r|r|}
\hline
\multicolumn{1}{|c|}{\textbf{Software}} &
\multicolumn{1}{|c|}{\textbf{Version}} &
\multicolumn{1}{|c|}{\textbf{Year}} &
\multicolumn{1}{c|}{\textbf{SLOC}} &
\multicolumn{1}{c|}{\textbf{E}} &
\multicolumn{1}{c|}{\textbf{D}} &
\multicolumn{1}{c|}{\textbf{P}} &
\multicolumn{1}{c|}{\textbf{FP}} &
\multicolumn{1}{c|}{\textbf{CC}} &
\multicolumn{1}{c|}{\textbf{CR}} &
\multicolumn{1}{c|}{\textbf{MI}}\\
\hline\hline
\texttt{Snort} & 2.9.8.2 & 2016         & 46,526 & 135.30 & 16.14 & 8.38 & 494.24 & 5.99 & 10.32 & 81.26\\
\hline
\texttt{Bash} & 4.4 rc-1 & 2016         & 160,890 & 497.81 & 26.47 & 18.81 & 2,265.35 & 12.6 & 17.08 & 35.61\\
\hline
\texttt{Apache} & 2.4.19 & 2016     & 280,051 & 890.86 & 33.03 & 26.97 & 4,520.10 & 5.45 & 23.42 & 62.58\\
\hline
\texttt{IPtables} & 1.6.0 & 2015     & 319,173 & 1,021.97 & 34.80 & 29.37 & 3,322.05 & 4.35& 27.33 & 49.98\\
\hline
\texttt{Git} & 2.8 & 2016             & 378,246 & 1,221.45 & 37.24 & 32.80 & 4,996.44 & 5.21 & 12.15 & 60.78\\
\hline
\texttt{Octave} & 4.0.1 & 2016      & 604,398 & 1,998.02 & 44.89 & 44.51 & 11,365.09 & 5.73& 27.69 & 41.60\\
\hline
\texttt{ClamAV} & 0.99.1 & 2016     & 714,085 & 2,380.39 & 47.98 & 49.61 & 10,669.97 & 6.36 & 33.57 & 63.01\\
\hline
\texttt{Cocos2d-x} & 3.10 & 2016 & 851,350 & 2,863.02 & 51.47 & 55.63 & 16,566.78 & 3.47& 17.55 & 68.18\\
\hline
\texttt{gcc} & 5.3 & 2015            & 6,378,290 & 2,3721.97 & 114.95 & 206.37 & 90,278.41 & 3.56 & 31.24 & 64.08\\
\hline
\end{tabular}
\caption{Software metrics for various open source projects.
\textbf{E}: COCOMO effort; \textbf{D}: COCOMO development time;
\textbf{P}: COCOMO team size; \textbf{FP}: function points;
\textbf{CC}: cyclomatic complexity; \textbf{CR}: comment-to-code ratio;
\textbf{MI}: maintainability index.
}
\label{table:metricsSW}
\end{table*}

\subsection{Comparison with Regular Software}
\label{sec:comparison}

In order to better appreciate the significance of the figures presented
so far, we next discuss how they compare to those
of a selected number of open source projects.
We selected 9 software projects:
3 security products (the \texttt{IPTables}
firewall, the \texttt{Snort} IDS, and the \texttt{ClamAV} antivirus);
a compiler (\texttt{gcc}); a web server (\texttt{Apache}); a version
control tool (\texttt{Git}); a numerical computation suite
(\texttt{Octave}); a graphic engine (\texttt{Cocos2d-x}); and a
Unix shell (\texttt{Bash}). The source code was downloaded from the
web page of each project. For each project we then computed
the metrics discussed above for malware. As in the case
of malware, we use the COCOMO coefficients for organic
projects. The results are shown in Table~\ref{table:metricsSW}
in increasing order of SLOC count.

The first natural comparison refers to the size of the source
code. Various malware samples from 2007 on (e.g. Zeus, 
KINS, Pony2, or SpyNet) have SLOC counts larger than those
of \texttt{Snort} and \texttt{Bash}. This automatically
translates, according to the COCOMO model, into similar or
greater development costs. The comparison of function point counts is
alike, with cases such as Rovnix and KINS having an FPC
greater than 1000, or SpyNet, with an FPC comparable
to that of \texttt{Bash}. In general, only complex malware
projects are comparable in size and effort to these two
software projects, and they are still far away from the
remaining ones.

\begin{figure}[t]
    \centering
        \includegraphics[width=1.05\columnwidth]{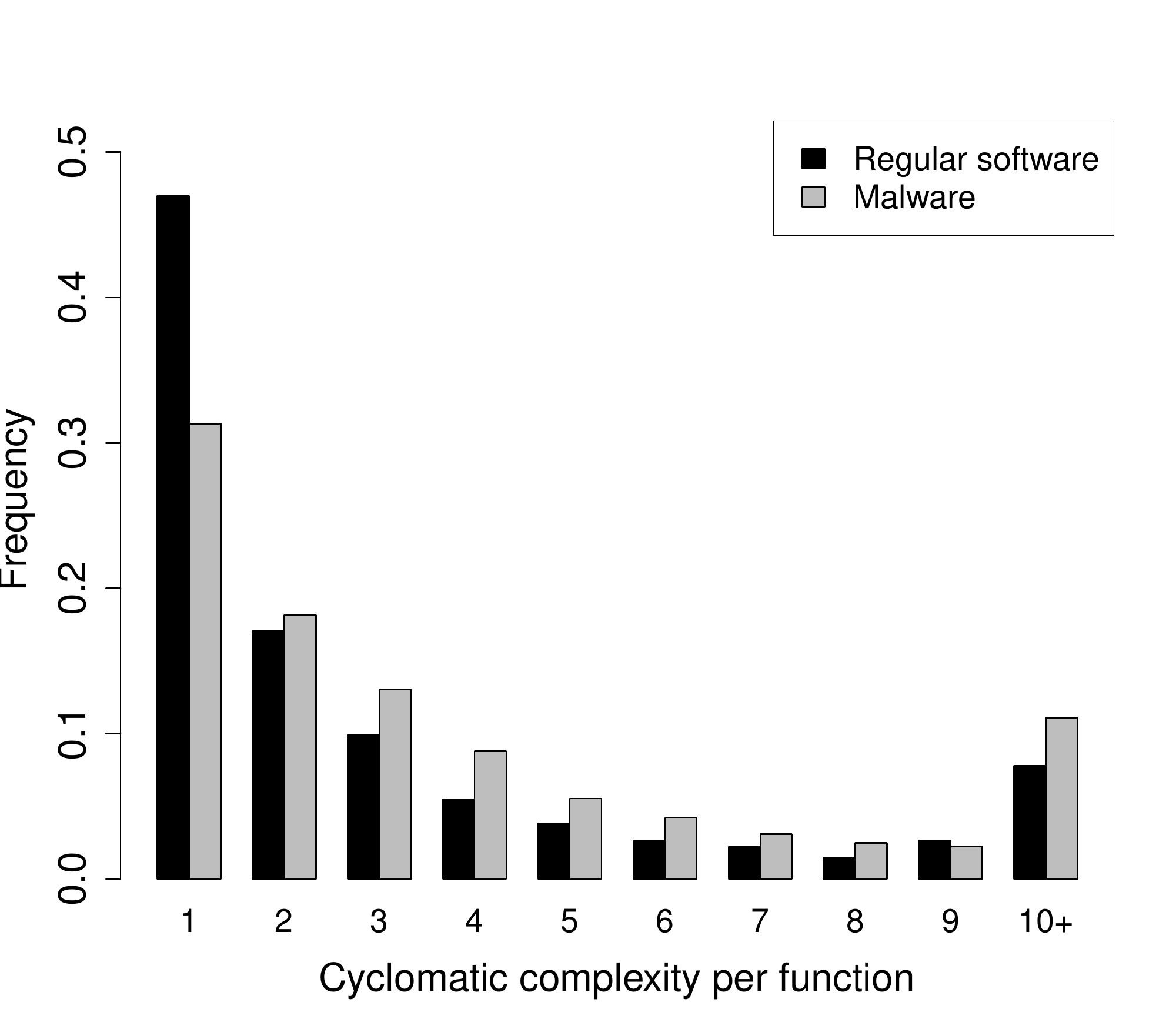}
        \caption{Histograms of the cyclomatic complexity values
        computed at the function level for both malware and regular
        software samples.}
        \label{fig:cyclo_mw_function}
\end{figure}

In terms of comments-to-code ratio, the figures are very
similar and there is no noticeable difference. This seems
to be the case for the cyclomatic complexity, too. To
further investigate this point, we computed the cyclomatic
complexities at the function level; i.e., for all functions
of all samples in both datasets. The histograms of the
obtained values are shown in Figure~\ref{fig:cyclo_mw_function}.
Both distributions are very similar, with a clear positive
skewness. A Chi-squared and two-sample Kolgomorov-Smirnov
tests corroborate their similarity for a significance level of
$\alpha = 0.05$.

Regarding the maintainability index, no 
malware sample in our dataset shows an $MI$ value higher than the
highest for regular software--\texttt{Snort}, with
$MI=81.26$. However, Figure~\ref{fig:maintainabilityMalware} 
shows that most $MI$ values for malware source code fall within the 
second and third quartiles, which also holds for traditional software. 
Two notable exceptions are Cairuh and the Fragus exploit kit, which
exhibit surprisingly low values (29.99 and 14.1, respectively).

\section{Source Code Reuse}
\label{sec:code_sharing}

This section presents the analysis of malware source code reuse in our dataset. 
Section~\ref{sec:cloneDetection} first introduces the two techniques we use 
for clone detection. 
We present the clone detection results in Section~\ref{sec:cloneResults}. 
Finally, Section~\ref{sec:cloneAnalysis} analyzes some of the clones found.

\subsection{Detecting Code Clones}
\label{sec:cloneDetection}

One challenge to detect code clones in our dataset is the diversity 
of programming languages used by the samples (Figure~\ref{fig:langs}).
Since samples written in C/C++ and Assembly constitute 92\% of our dataset
(115 projects contain at least one file fully written in C/C++ and 304 
projects contain at least one file fully written in Assembly),
we need clone detection techniques that can at least cover these two languages.
To achieve this goal, we use two code detection techniques detailed next.

\begin{figure}[h]
\begin{minipage}[t]{0.4\linewidth}
\begin{lstlisting}[language=c++]
int InfectExes(void){
WIN32_FIND_DATA d32;
char MyFile[256];
GetFileName(MyFile,sizeof(MyFile));
\end{lstlisting}
\end{minipage}%
\hfill\hrule
\begin{minipage}[t]{0.45\linewidth}
\begin{lstlisting}[language=c++]
int InfectFiles(void){
WIN32_FIND_DATA w32;
char FileName[256];
GetFileName(FileName,sizeof(FileName));
\end{lstlisting}
\end{minipage}%
\caption{Two cloned code fragments sharing identical syntactic structure but with different names for method and variables.}
\label{fig:similar_code_fragment}
\end{figure}

\paragraph{Deckard.}
Our first clone detection technique uses Deckard~\cite{jiang2007deckard}, 
a tool for detecting source code clones
that was specifically designed to scale to large code bases
such as the Java JDK and the Linux kernel,
which comprise thousands of files. 
Deckard computes an Abstract Syntax Tree (AST) for each input source file.
For each AST tree it produces a set of vectors of fixed length, 
each representing the structure of the AST subtree rooted at a specific node.
These vectors are then clustered and each output cluster 
comprises of a set of similar ASTs, each a clone of the others.
One advantage of AST-based clone detection techniques is that they 
can detect code clones with the same structure, 
despite some code changes such as variable renaming or different 
values assigned to a variable. 
Figure~\ref{fig:similar_code_fragment} shows an example of a code clone 
detected by Deckard despite changes in the names of the function, 
function parameters, and function variables.
On the other hand, they can have high false positives 
since code snippets with the same AST structure may not necessarily be clones.

To limit the false positives, Deckard allows to specify the 
\emph{minimum clone size} (as a number of AST tokens). 
This enables to remove short code sequences that appear in 
multiple samples simply because they are commonly used, 
but may not imply that the code was copied from one project into another.
For example, sequences of C/C++ \texttt{\#include} directives and 
loop statements, e.g., \texttt{for (i=0; i<n; i++)}, are not real clones. 
Deckard allows the user to set two additional parameters, 
the \emph{stride} that controls the size of the sliding windows used during 
the serialization of ASTs, and 
the \emph{clone similarity} used to determine if two code fragments are clones.
In our experiments we tried different settings for these parameters. 
We obtained best results using minimum clone size of 100, 
stride of 2, and 1.0 similarity.

By default, Deckard offers support for the following languages: 
C/C++, Java, PHP, and dot.
It can also support other languages if a grammar is available.
Since our dataset contains a diversity of Assembly instruction sets (PPC, x86)
and syntax specifications (Intel, AT\&T),
we would need to generate a grammar for each instruction set and syntax. 
That would require a significant effort to support Assembly samples 
(in some cases with a low return given the small number of samples for 
some combinations). 
Thus, we only use Deckard for finding clones among samples 
developed using C/C++.
We apply Deckard on all projects of the same language (C/C++ or Assembly) 
simultaneously, which leverages Deckard's design for efficiency. 

\paragraph{\Diff.}
Our second clone detection technique compares
two source code files using the Ratcliff-Obershelp algorithm 
for string similarity~\cite{ratcliff1988pattern}. 
This technique measures how similar 
two sequences of characters are by computing the ratio between the matching 
characters and the total number of characters in the two sequences. The 
algorithm outputs matching blocks of characters containing the longest 
common subsequence (LCS) and characters neighboring the LCS that are 
similar in both strings. 
We consider a code clone every matching block 
satisfying a minimum length.
We experimented with different minimum length values, 
achieving best results with a minimum length of 10 SLOC for Assembly and 
5 SLOC for C/C++.
The higher threshold for Assembly is due to its lower abstraction compared 
to C/C++.

Since this technique operates on two input files, 
for each pair of samples we compare every pair of source code files, 
one file from each sample. 
To avoid missing clones because of simple changes to the copied code 
we preprocess the files using these rules:
remove blank spaces, tabs, and newline characters; 
convert to lower case; and 
translate the character set to UTF-8. 

The main advantages of this technique are its simplicity, 
that it can handle any text-based language, 
and very low false positives. 
The main disadvantages are potentially high false 
negatives and low efficiency. 
False negatives can happen because this technique only detects 
reuse of identical code fragments; 
it will not detect a clone if the code is modified
(e.g., variable renaming).
Low efficiency is due to the quadratic number of comparisons needed.

\begin{table*}[t!]
\centering
\begin{tabular}{|l|l|r|r|r|r|r|r|r|r|r|r|}
\hline
\multirow{2}{*}{\textbf{Language}} &
\multirow{2}{*}{\textbf{Detection technique}} &
\multirow{2}{*}{\textbf{Clones}} &
\multirow{2}{*}{\textbf{TPs}} &
\multirow{2}{*}{\textbf{FPs}} &
\multicolumn{2}{c|}{\textbf{Runtime}} &
\multicolumn{5}{c|}{\textbf{Clone size (SLOC)}}\\
\cline{6-12}
& & & & & 
\multicolumn{1}{c|}{\textbf{All (h)}} &
\multicolumn{1}{c|}{\textbf{Pair (s)}} &
\multicolumn{1}{c|}{\textbf{Average}} &
\multicolumn{1}{c|}{\textbf{Std. Dev.}} &
\multicolumn{1}{c|}{\textbf{Median}} &
\multicolumn{1}{c|}{\textbf{Min.}} &
\multicolumn{1}{c|}{\textbf{Max.}} \\
\hline\hline
C/C++ &
Deckard &
7,655 &
984 &
6,671&
1.4 &
0.8 &
112.3 &
1,441.2 &
17 &
7 &
22,658 \\
\hline
C/C++ &
\Diffshort (raw) &
1,040 & 
973 &
67 &
197.7 &
107.5 &
25.9 &
69.4 &
7 &
5 &
1,157 \\
\hline
C/C++ &
\Diffshort (with normalization) &
959 &
904 &
55 &
209.3 &
115.0 &
52.7 &
762.9 &
7 &
5 &
22,709 \\
\hline
\hline
\hline
Assembly &
\Diffshort (raw) &
974 &
972 &
2 &
97.8 &
7.6 &
50.1 &
100.1  &
19 &
10 &
1,084 \\
\hline
Assembly &
\Diffshort (with normalization) &
704 &
703 &
1 &
101.0 &
7.9 &
58.6 &
102.8 &
21 &
10 &
1,084 \\
\hline
\end{tabular}
\caption{Clone detection results for C/C++ and Assembly projects using 
the two detection techniques.
It first shows the number of code clones detected and their split into  
true positives and false positives. 
Next, it shows the total runtime in hours and the average runtime for each pair
of projects in seconds.
Finally, it shows clone size statistics.
}
\label{table:pairwise}
\end{table*}

\ignore{
\begin{table*}[t!]
\centering
\label{times}
\begin{tabular}{|l|l|l|l|l}
\cline{1-4}
\multicolumn{2}{|l|}{\textbf{Technique}} & 
\textbf{Avg. time per pair (s)} & 
\textbf{Total time (s)} &  
\\ 
\cline{1-4}
\multirow{4}{*}{\textbf{\Diff}} & 
C/C++ (raw)  &
107.48 &
704556.67 &  
\\ \cline{2-4}& 
C/C++ (with normalization) & 
114.97 &
753672.36 &  
\\ \cline{2-4}&
Assembly (raw) &
7.65 &
352202.57 & 
\\ \cline{2-4} &
 Assembly (with normalization) &
 7.89 &
 363594.78 & 
 \\ \cline{1-4}
\textbf{Deckard}  & 
C/C++&
0.788 &
5170.78 & 
\\ \cline{1-4}
\end{tabular}
\caption{Runtime measurements for each technique.}
\end{table*}
}

\subsection{Clone Detection Results}
\label{sec:cloneResults}

This section presents the clone detection results using Deckard and 
the pairwise comparison technique.
We manually examine the clones detected by both techniques to determine 
whether they are true positives or false positives.
During our initial manual analysis we found that a large 
number of clones were different instances of the same cases. 
To speed up the manual analysis of the more than 10K detected clones, 
we use a clustering approach based on 
regular expressions and fuzzy hashing~\cite{ssdeep} to automatically group 
nearly identical clones. 
The analyst then labels each cluster, which considerable 
speeds up the manual analysis since the number of clusters is 
nearly two orders of magnitude smaller than the number of clones. 

Table~\ref{table:pairwise} summarizes the code clone detection results. 
For each language and detection technique, it shows 
the number of detected clones,
the split of those clones into true and false positives, 
the total and per-pair runtime, 
and statistics on the SLOC size of the detected clones.

The C/C++ results show that Deckard detects 7,655 clones compared to 
959--1,040 using the \diff technique. 
However, of the 7,655 Deckard clones, 
87\% are false positives, 
compared to 6.4\% (raw) and 5.7\% (normalized) using \diff.
The very high false positive rate of Deckard is due to its 
AST representation, which ignores type information and constant values. 
For example, an array definition like
{\tt static unsigned long SP1[64] = \{ 0x01010400L, $\dots$ \} } is  
considered by Deckard a clone of
{\tt static unsigned char PADDING[64] = \{0x80, $\dots$ \}},
even if both arrays are of different type and are initialized with 
different values.
As another example, the function invocation 
{\tt  CopyFile(wormpath, "gratis.mp3.exe", 0)} is 
considered a clone of
{\tt  add\_entry(TABLE\_KILLER\_FD, "{\textbackslash}x0D{\textbackslash}x44{\textbackslash}x46{\textbackslash}x22", 4)}.

\paragraph{Clone lengths.}
The average clone size using Deckard is 112.3 SLOC, 
52.7 using normalized \diff, and 25.9 using raw \diff. 
Thus, while the number of TPs is similar using Deckard and raw \diff, 
Deckard is able to find longer (i.e., higher quality) clones. 
Surprisingly, the number of TPs found by the raw \diff 
is higher than those found by the normalized \diff. 
This happens because the raw \diff breaks longer 
clones into multiple smaller clones, 
which increases the number of detected clones, but produces shorter 
(i.e., lower quality) clones.
Thus, normalization helps to find larger clones. 
For example, in the C/C++ code, normalization allowed us to discover a 
large true clone consisting of 22K SLOC 
(detailed in Section~\ref{sec:cloneAnalysis}).

\begin{figure*}[t]
\centering
\begin{subfigure}{0.9\columnwidth}

\includegraphics[width=0.9\columnwidth]{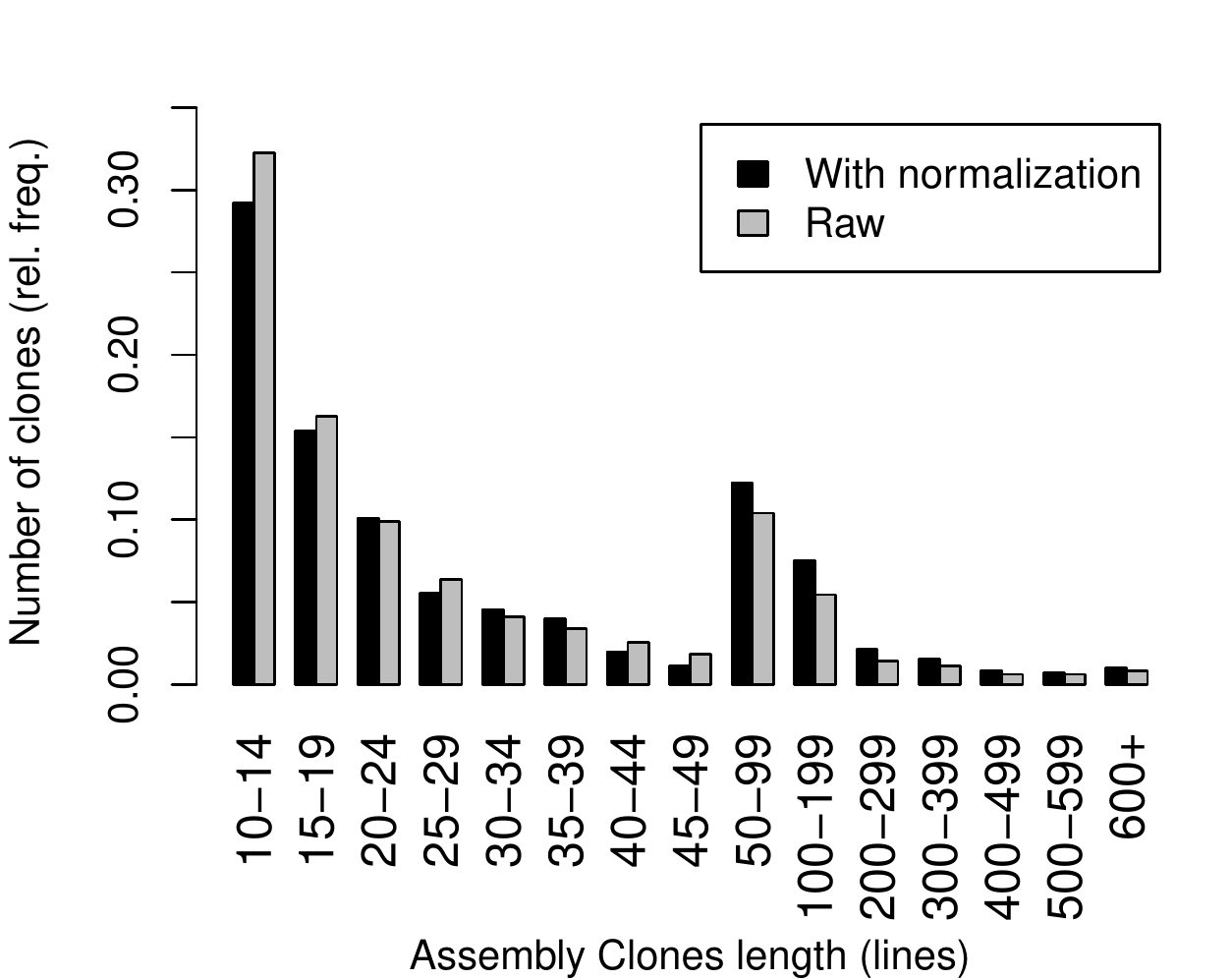}%
\caption{}
\label{fig:distASM}%
\end{subfigure}
\begin{subfigure}{0.9\columnwidth}
\includegraphics[width=0.9\columnwidth]{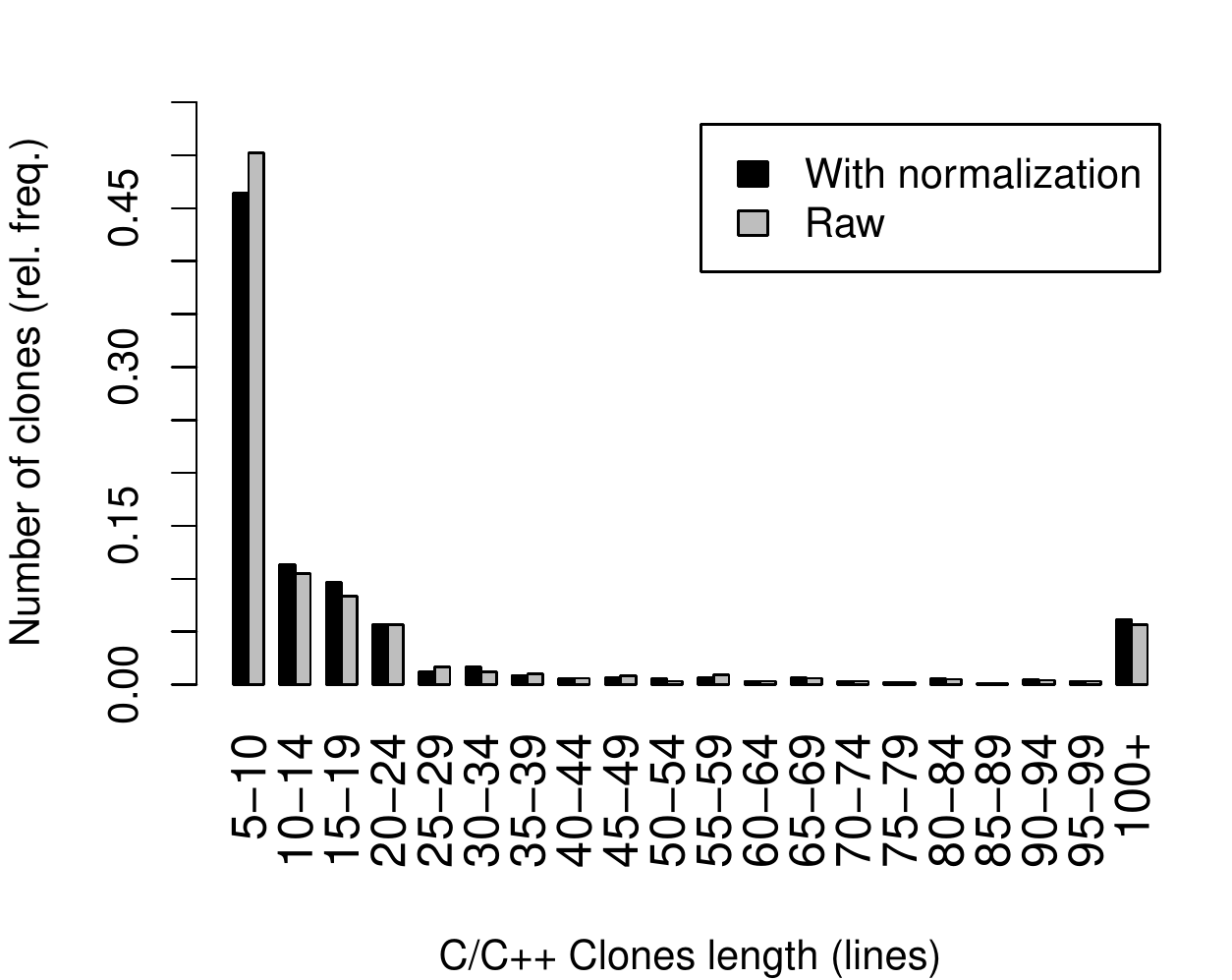}%
\caption{}
\label{fig:distC}%
\end{subfigure}\hfill

\caption{Distribution of code clone sizes for Assembly and C/C++ languages}
\label{fig:lengthdist}
\end{figure*}

Figure \ref{fig:lengthdist} shows the distributions of length values for 
raw and normalized clones in both languages. 
In both distributions the number of clones becomes smaller as the
size grows, which translates into positively skewed distributions. 
Noteworthy exceptions of this trend are clones in the range of 
50-99 and 100-199 lines in Assembly code, and
also clones larger than 100 lines in C/C++. 
These peaks are related to the nature
of the detected clones, discussed in Section~\ref{sec:cloneAnalysis}.

Small clones (i.e., shorter than 20 lines) are fairly common in both 
C/C++ and Assembly. Manual inspection revealed different explanations for
each language. In the case of Assembly, such small cloned fragments are
usually related to control flow structures such as loops, which employ the
same sequence of instructions regardless of the actual data values. In
addition, we also observed reuse of the names used to label Assembly code
segments. In the case of C/C++ projects, we found that small clones are
generally associated with preprocessor directives such as \texttt{\#typedef},
\texttt{\#define}, and \texttt{\#include}. These short clones also include
sequences of instructions to initialize data structures (e.g., arrays),
and generic sequences often found at the end of functions that release
allocated variables before returning a value. These clones are often exact
copies of each other and are common in malware samples from the same family.

On the other hand, clones larger than 20 lines represent less than
50\% of the detected clones in both languages. In particular, 350
assembly clones are larger than 20 lines. The number of Assembly clones is
also notably smaller than the total number of Assembly source code files in
our dataset, which is close to 800. Comparing the average lengths of 
Assembly source code files and large clones provides similar results. Large
Assembly clones are 102.87 SLOC on average, while Assembly source files in
the dataset are 498.18 SLOC. In the case of C/C++ clones, the figures are
comparable. We identified 450 C/C++ clones out of 984 that are larger than
20 lines. The total number of C/C++ files in the dataset is 2,981, which
almost doubles the total number of clones found. The average length of
large C/C++ clones depends greatly on the normalization process: it is
just 102.01 SLOC without normalization and 231.28 SLOC after normalization.

We analyzed the size of clones found in samples belonging to the same
family or developed by the same author. To do so, we selected 4 pairs of
projects developed in C/C++ and Assembly for which we had ground truth, i.e.,
they are known to share authorship or are variants of the same family.
Specifically, we selected two banking trojans (Zeus \& Kins) and two worms
(Cairuh.A \& Hunatch.a) for the C/C++ case; and two pairs of viruses written
in Assembly: (methaphor.1.d \& Simile) and (EfishNC \& JunkMail). The average
clone sizes for the C/C++ couples are 57.40 lines (37 clones) for the banking
trojans and 13.86 lines (60 clones) for the worms. In the case of the Assembly
samples, the average clone lengths are 179.72 lines for methaphor.1.d \& Simile
(54 clones) and 47.06 lines for EfishNC \& JunkMail (30 clones).

\begin{figure}
\includegraphics[width=\columnwidth]{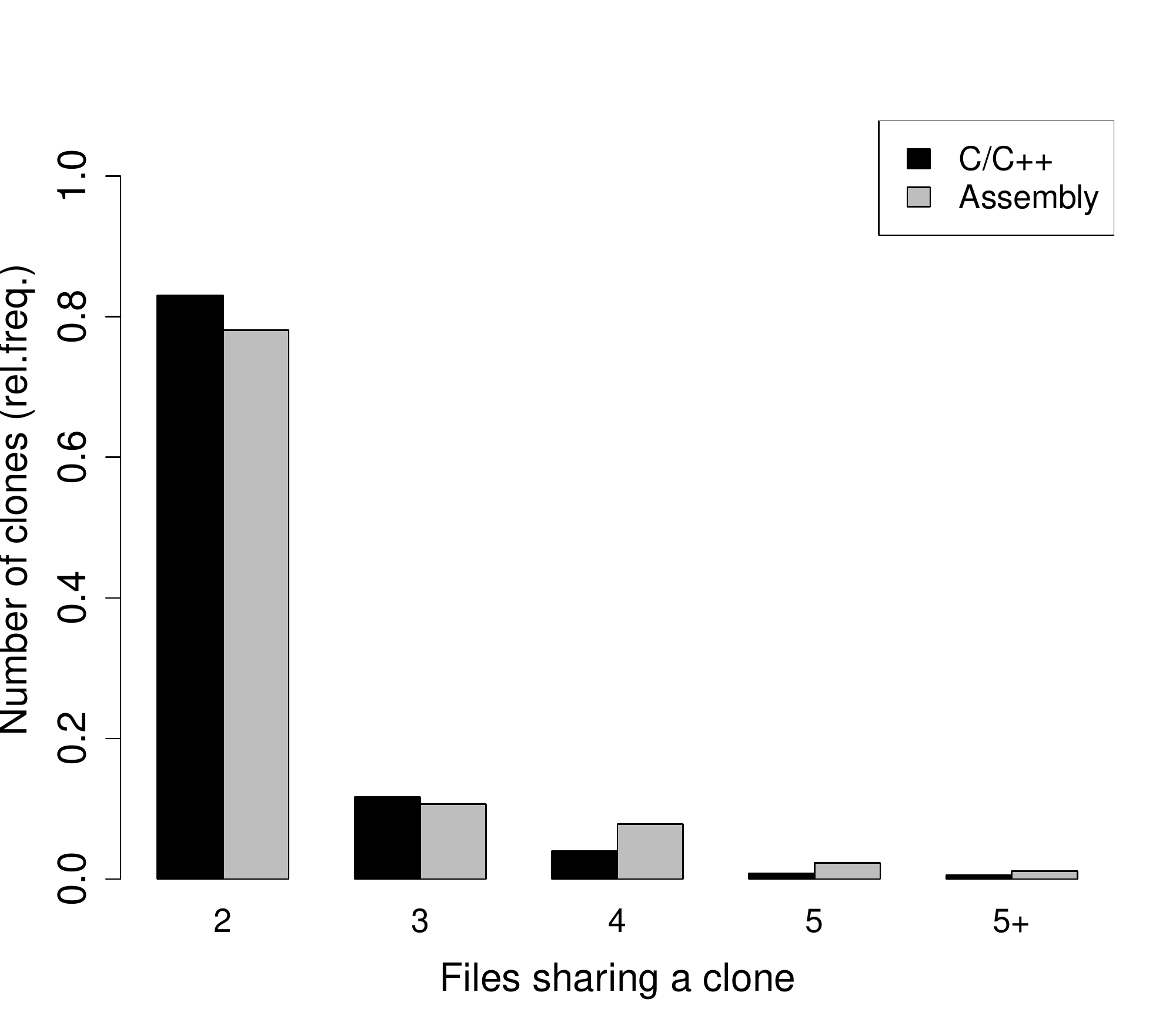}
\caption{Distribution of clones in 2 or more files.}
\label{fig:cloneDistribution}
\end{figure}

\paragraph{Clone file distribution.}
Figure \ref{fig:cloneDistribution} shows the distribution of the
number of files in which a clone appears. As it can be seen, in approximately
80\% of the cases clones are found in just two files. The number of clones
appearing in 3 or more files decreases considerably for both languages. In the
case of C/C++, the fraction of clones appearing in 3, 4, 5, and more than 5 files
is 0.11, 0.04, 0.008, and 0.005 respectively. The pattern for Assembly clones is
similar, though clones are slightly more widespread as they appear in more than
3 files more often than C/C++ clones.

\paragraph{Runtime.}
Deckard is almost two orders of magnitude faster than the \diff technique, 
finding clones across all 115 C/C++ samples in 1.4 hours, 
compared to 8 days for the \diff.
Such efficiency difference is due to Deckard parsing each file only once 
and to its clustering.
We observe that the \diff on Assembly run much faster than on C/C++. 
This is due to the C/C++ projects containing more, and longer, files.

To conclude, our clone detection results show a significant amount 
of code reuse despite the relatively small number of samples, 
e.g., 984 clones of 112 SLOC on average across 115 C/C++ projects. 
We detail the type of clones found in Section~\ref{sec:cloneAnalysis}.
This constitutes an evidence that malware authors 
copy functionalities useful to their goals that may appear in 
leaked malware source code (or malware they have purchased).
Of the two techniques evaluated, Deckard runs very fast and finds more 
and longer clones, but produces very high false positives.
On the other hand, the \diff technique produces much lower 
(but a non-negligible 6\%) false positives, 
but runs two orders of magnitude slower.

\subsection{Clone Analysis}
\label{sec:cloneAnalysis}

We next discuss the purpose of the reused code fragments we found. 
Through manual analysis, we classified clones into four main groups 
according to the functionality they provide. For additional details, 
Tables \ref{table:clones_c} and \ref{table:clones_assembly} summarize 
the main features of a selection of representative 
clones of these categories we found for both C/C++ and Assembly.

\paragraph{Operational data structures and functions.}
One large group of code clones consists of libraries of data structures and the associated
functions to manipulate system and networking artifacts, such as executable file formats
(PE and ELF) and communication protocols (TCP, HTTP) and services (SMTP, DNS). We also observe 
a number of clones consisting of headers for several API functions needed to interact
with the Windows kernel, such as the 3,054 lines long clone shared by W32.Remhead and
W32.Rovnix.

\paragraph{Core malware artifacts.}
The second category of clones consists of code that implements properly malicious
capabilities, such as infection, spreading, or actions on the victim. For instance,
the W32.Dopebot botnet contains shellcode to exploit the CVE-2003-0533 vulnerability, and
the same shellcode is found in the W32.Sasser worm. Another good example of this practice  is
the network sniffer shared by W32.NullBot and W32.LoexBot.

\paragraph{Data clones.}
Some of the clones are not code, 
but rather data structures that appear in multiple samples. 
An example is the array of frequent passwords present in both 
W32.Rbot and W32.LoexBot. 
Another example is the list of strings found in W32.Hunatchab and W32.Branko, 
containing the process names associated to different commercial 
AV (antivirus) software, which both bots try to disable.
Furthermore, some samples also share strings containing IP addresses, 
for example the Sasser worm and the Dopebot botnet.

\paragraph{Anti-analysis capabilities.}
One of the most noticeable examples of this is the packer included in the W32.Cairuh worm,
which is shared by the W32.Hexbot botnet. Its size is 22,709 lines and it is the biggest
clone we found in our dataset. Another remarkable example is the metamorphic engine shared
by the Simile and Metaphor.1d viruses, consisting of more than 10,900 lines of assembly code.
Other examples of reused anti-analysis modules can be found in W32.Antares and W32.Vampiro,
which share the same polymorphic engine, and also in W95.Babyloni, and W32.Ramlide, which
share the same packing engine. Finally, we also found a number of reused instances of
code to kill running AV processes, such as the clone found in Hunatchab.c and
Branko.c

\begin{table}[]
\centering
\begin{tabular}{c|c|c|c|l|}
\cline{2-5}
\multicolumn{1}{l|}{}                 & \multirow{2}{*}{\textbf{Assembly}} & \multirow{2}{*}{\textbf{C/C++}} & \multicolumn{2}{c|}{\textbf{Avg. length (SLOC)}} \\ \cline{4-5} 
\multicolumn{1}{l|}{}                 &                                    &                                 & \textbf{Assembly}        & \textbf{C/C++}        \\ \hline
\multicolumn{1}{|c|}{\textbf{Type A}} & 22                                 & 42                              & 364.86                   & 193.83                \\ \hline
\multicolumn{1}{|c|}{\textbf{Type B}} & 55                                 & 41                              & 238.94                   & 151.70                \\ \hline
\multicolumn{1}{|c|}{\textbf{Type C}} & 15                                 & 8                               & 101.6                    & 121.62                \\ \hline
\multicolumn{1}{|c|}{\textbf{Type D}} & 8                                  & 9                               & 613.62                   & 2587.33                 \\ \hline
\end{tabular}
\caption{Clone types frequencies for Assembly and C/C++ by types. 
\textbf{Type A: }Operational data structures and functions, \textbf{Type B: }Core malware artifacts, 
\textbf{Type C: }Data clones, \textbf{Type D: }Anti-analysis capabilities.}
\label{table:clone-type-dist}
\end{table}

In order to estimate the number of clones for each category, we randomly 
sampled the set of found clones and selected 100 for each language. The 200
clones were then manually labeled according to the four semantic categories
described above. Table \ref{table:clone-type-dist} shows the distribution of
clones together with their average length. As it can be seen, most of the
cases belong to types A (operational data structures and functions) and B
(core malware artifacts). In the case of Assembly, both categories amount for 
84\% of all clones, while in the case of C/C++ core malware artefacts
alone constitute 55\% of the clones. In both cases, data clones and
anti-analysis capabilities are considerably less frequent.

With respect to their lengths, Type D assembly clones are noticeably larger
than clones in other categories. This is due to the presence of polymorphic
and packing engines in this category, which are relatively complex code
samples. Contrarily, data clones (Type C) are generally shorter, which is
reasonably given their nature. In general, Assembly clones are bigger than
their C/C++ counterpart, which is in line with the general results described
above.

The data in Table \ref{table:clone-type-dist} suggests 
that clone size highly depends on the nature of shared 
features. This is especially evident for those clones labeled 
as type C. In addition, the results reveal that the inclusion 
of evasion and anti-analysis capabilities has a noticeable impact
in the size of malicious codebases.

Last but not least, we observed that in most cases code reuse usually takes place in short
time spans, i.e., the samples sharing a particular fragment of code have been developed within
1-4 years of each other. This could evidence that the same author has participated in the
development of such samples, or else that collaborating groups share previously developed
artifacts that can be easily reused.

\begin{table*}[]
\centering

\begin{tabular}{|c|c|c|c|}
\hline
\textbf{Length} & \textbf{Samples} & \textbf{Description} & \textbf{Category} \\ \hline
22709 & \begin{tabular}[c]{@{}c@{}}W32.Cairuh.A (Worm, 2009)\\ W32.Simile (Worm, 2009)\\ W32.HexBot2 (Bot, 2009)\end{tabular} & \begin{tabular}[c]{@{}c@{}}Array containing a raw dump of an executable packing tool used \\ after the compilation of the main binary\end{tabular} & Anti-analysis capabilities \\ \hline
3054 & \begin{tabular}[c]{@{}c@{}}W32.Remhead (Trojan,2004)\\ W32.Rovnix (Virus,2014)\\ W32.Carberp (Virus,2013)\end{tabular} & \begin{tabular}[c]{@{}c@{}}Define several data structures used for interacting with the \\ NT Kernel through its Native API.\end{tabular} & \begin{tabular}[c]{@{}c@{}}Operational data structures\\ and functions\end{tabular} \\ \hline
2546 & \begin{tabular}[c]{@{}c@{}}W32.MyDoom (Worm, 2004)\\ W32.HellBot (Bot, 2005)\end{tabular} & Share the code for deploying a SMTP relay and a fake DNS server. & Core malware artifacts \\ \hline
1323 & \begin{tabular}[c]{@{}c@{}}W32.NullBot (Bot, 2006)\\ W32.LoexBot (Bot, 2008)\end{tabular} & Includes hardcoded IRC commands used to send instruction to infected clients & Data clones \\ \hline
328 & \begin{tabular}[c]{@{}c@{}}W32.Dopebot.A (Bot, 2004)\\ W32.Dopebot.B (Bot, 2004)\\ W32.Sasser (Worm, 2004)\end{tabular} & Shellcode employed for exploiting, the CVE-2003-0533 vulnerability \cite{vuln} & Core malware artifacts \\ \hline
\end{tabular}
\caption{Examples of code clones found in C/C++.}
\label{table:clones_c}
\end{table*}

\begin{table*}[]
\centering
\begin{tabular}{|c|c|c|c|}
\hline
\textbf{Length} & \textbf{Samples}& \textbf{Description} & \textbf{Category}\\ \hline

10917& \begin{tabular}[c]{@{}c@{}}W32.Metaph0r.1d (Virus, 2002)\\ W32.Simile (Virus, 2002)\end{tabular} & \begin{tabular}[c]{@{}c@{}}These samples contain a complete metamorphic engine \\ coded in pure x86 assembly.\end{tabular}& Anti-analysis capabilities \\ \hline

1283& \begin{tabular}[c]{@{}c@{}}W32.EfishNC (Virus,2002)\\ W32.Junkmail (Virus,2003)\end{tabular}& \begin{tabular}[c]{@{}c@{}}Both declare the same structs and function headers for \\ infection and spreading through email\end{tabular}& Core malware artifacts\\ \hline

233 & \begin{tabular}[c]{@{}c@{}}Lin32.Tahorg (Virus, 2003)\\ Lin32.GripB (Virus, 2005)\end{tabular} & \begin{tabular}[c]{@{}c@{}}Share structures and routines for reading and modifying ELF files. \\ Includes a list of offsets for the different sections in the EFL header.\end{tabular}& \begin{tabular}[c]{@{}c@{}}Operational data structures \\ and functions.\end{tabular} \\ \hline

1009& \begin{tabular}[c]{@{}c@{}}W32.Relock (Virus,2007)\\ W32.Mimix (Virus,2008)\\ W32.Impute (Virus, 2013)\end{tabular} & \begin{tabular}[c]{@{}c@{}}These samples share an assembly implementation of Marsenne Twister PRNG.\\  W32.Relock was the first malware piece using the \textit{virtual code} \\ obfuscation technique \cite{virtual_code} which is based in memory reallocation.\end{tabular} & Anti-analysis capabilities \\ \hline

100  & \begin{tabular}[c]{@{}c@{}}Gemini (Virus,2003)\\ EfisNC (Virus,2008)\\ JunkMail (Virus,2013)\end{tabular}           & \begin{tabular}[c]{@{}c@{}}Contains offsets pointing to several functions within a MZ \\ (DOS Executable files) manipulation library\end{tabular}& Data clones\\ \hline

\end{tabular}
\caption{Examples of code clones found in Assembly.}
\label{table:clones_assembly}
\end{table*}

\subsection{Code Sharing with Benign Source Code}
\label{sec:malbensharing}
We also explored if code cloning between malicious and benign source 
happens to the same extent as it does among malware samples. For this purpose, 
we analyzed the set of major open source projects used in Section
\ref{sec:comparison}  and extended this set adding the Bitcoin cryptocurrency
and Linux kernel source code master branches.

We followed a similar approach as for the main cloning experiment.
However we decided to lean exclusively on Deckard since it is faster,
especially when dealing with large codebases. We ran Deckard ten times, one
time per project, combining the open source project with the whole malicious
source code dataset each time. Then, we processed the output as outlined in
Section \ref{sec:cloneDetection}. Despite the high FP ratios obtained, in the
experiment we found code cloning cases in 4 out of 10 source code projects.
Snort, Iptables, Bash, Apache, Cocos2d and the Bitcoin projects do not share
any source code snippet with any of the samples included in our dataset.
We did find up to $210$ relevant code clones (larger than 5 lines) in
\texttt{gcc}, the Linux kernel, \texttt{Git}, and \texttt{clamAV}. 
Surprisingly, all the cloned source clones found in \texttt{gcc},
\texttt{clamAV}, and the Linux kernel are part of the \texttt{Zlib} compression
library. In particular, the cloned fragments appear in a C header
(\texttt{defutil.h}) and 3 C source files (\texttt{infbak.h},
\texttt{inflate64.c}, and \texttt{inflate.c}) in the open source project. In
the malicious source code dataset, the same fragments are contained in the
files \texttt{deflate.h}, \texttt{infback.c}, \texttt{inflate.c}, and
\texttt{inftrees.c} included in the server source code of the XtremeRAT botnet.
Git shares a set of data structures with the samples w32.Rovnix and
w32.Carberp. The content of these data structures is used as padding in the
implementation of the SHA1 and MD5 hashing algorithms. The shared code is
located in the files \texttt{sha1.c} in the git source code tree and also in
the files \texttt{md5.c} and \texttt{md5.cpp} included in the code of Rovnix
and Carberp, respectively. The average size of the cloned fragments is 102
lines.

\section{Discussion}
\label{sec:discussion}

We next discuss some aspects of the suitability of our approach,
the potential limitations of our results, and draw some general
conclusions.

\paragraph{Suitability of our approach.}
Software metrics have a long-standing tradition in software engineering
and have been an important part of the discipline since its early days.
Still, they have been subject to much debate, largely
because of frequent misinterpretations (e.g., as performance
indicators) and misuse (e.g., to drive management)~\cite{Sommerville06}.
In this work, our use of certain software metrics pursues a
different goal, namely to quantify how different properties of
malware as a software artifact have evolved over time. Thus,
our focus here is not on the accuracy of the absolute values (e.g., effort
estimates given by COCOMO), 
but rather on the relative comparison of values between malware samples, 
as well as with benign programs, and 
the trends that the analysis suggests.

The use of comments as an efficient documentation method has 
been questioned by several experts. Among the stated reasons, 
it has been argued that often comments add redundant description 
of code functionality instead of clarifying design decisions and the 
underliying algorithmic workflow. However others authors defend 
that good quality comments are still valuable and necessary, 
specially in large collaborative projects \cite{acm_blog}.
The validity of the comments-to-code ratio 
nowadays could also be criticized, 
given the trend to develop source code using automatically 
generated documentation frameworks. 
This trend may have reduced over time the reliability of 
comments-to-code ratio as a maintainability metric. 
Nevertheless, during our analysis we did not find any samples, 
using such approaches, 
as the only delivered documentation with the (recent) samples,
are the comments written by the authors. 
Thus, comments seem to still play an important role in the development of 
malware.

As for the case of detecting code reuse, the techniques we used 
represent standard approaches to the problem. By using two different 
approaches, we obtain complementary and more robust results.
For example, we can use the \diff technique to analyze assembly samples 
not supported by Deckard, while Deckard's AST-based approach 
resists certain classes of evasion attacks, 
e.g, variable and function renaming, 
which affect the \diff technique.

\paragraph{Limitations.}
Our analysis may suffer from several limitations.
Perhaps the most salient is the reduced number of samples in our
dataset. However, as discussed in Section~\ref{sec:dataset}, obtaining
source code of malware is hard.
Still, we analyze 456 samples, 
which to the best of our knowledge is the largest dataset 
of malware source code analyzed in the literature.
While the exact coverage of our dataset cannot be known, 
we believe it is fairly representative in terms of different
types of malware.
It should also be noted that in our study, the sample concept
refers to a malware family. Thus, we are not only covering 456 binary
samples but a wider set of potential variants. The wide gap between the
number of binary samples found in the wild and the number of malware 
families has been previously discussed in the community.
A recent study \cite{lever2017lustrum} examined 23.9M samples
and classified them into 17.7K families (i.e., three orders of magnitude
smaller). While this phenomenon is due to different reasons, the most prominent one
is the use of polymorphism and other advanced obfuscation methods employed
by malware authors. We note that 428 out of 17.7K is a respectable 2.4\% coverage.

In particular, we believe the coverage of our dataset is 
enough to quantify and analyze the trends in malware evolution 
(size, development cost, complexity), 
but we do not attempt to analyze the evolution of malware code reuse.
Since we only have one (or a few) versions for each malware family 
and a limited number of families, our dataset may miss important 
instances of malware code reuse. 
Thus, we have focused on analyzing what type of code we observe being reused 
in our dataset.
As we collect more samples,
we should be able to obtain a more representative picture of the code 
sharing phenomenon in malware creation, 
going beyond the findings we have reported.

Another limitation is selection bias. 
Collection is particularly difficult for newest samples and 
more sophisticated samples 
(e.g., those used in targeted attacks) 
have not become publicly available. 
We believe those samples would emphasize the increasing 
complexity trends that we observe. 

Finally, even if the employed \diff code clone 
detection technique is very simple and has poor scalability, 
it has performed remarkably well in terms of false positives 
compared with Deckard, a more sophisticated tool
based on comparing syntactic structures. 
The large amount of false positives obtained with Deckard 
can be partially explained because of
the way in which malware writers reuse code. As discussed in
section \ref{sec:code_sharing}, cloned code fragments
are often core artifacts such as shellcodes or
obfuscation engines. Given the nature of these
artifacts, malware authors are forced to reuse them in a copy 
and paste fashion rather than rewriting some of their content. 
This makes very uncommon to find partial clones, consisting on
slightly modified code fragments. For this reason, and despite the 
great variety of code-clone detection techniques available in the 
literature~\cite{roy2007survey,sheneamer2016survey}, it is unclear whether
employing more sophisticated approaches might lead to finding 
significantly more clones when dealing with plain malware source code.

In addition, clone detection tools based on syntactic structures 
depend greatly on the set of selected features. 
In the case of Deckard, leaving out data types and literals definitely
contributes to achieving poorly accurate results, especially in our use
case which differs from standard use cases for this kind of tools.

Deckard could be improved in many ways in order to obtain more
precise results. Two natural ideas would be combining syntactic and semantic
features, and introducing a similarity metric after the clustering step.
However, in this paper we just aimed at comparing the performance of a 
naive approach (diff-based clone detection) against an already proposed tool,
and therefore we decided to use Deckard out of the box, leaving out any
improvement.

\paragraph{Main conclusions and open questions.}
In the last 40 years the complexity of malware, considered as a
software product, has increased considerably. We observe
increments of nearly one order of magnitude per decade in
aspects such as the number of source code files, 
source code lines, and function point counts.
This growth in size can be attributed to various
interconnected reasons. On the one hand, malicious code has progressively
adapted to the increasing complexity of the victim platforms they target.
Thus, as Operating Systems evolved to offer richer application
programming interfaces (API), malware authors rapidly leveraged them
to achieve their purposes. This translated into larger and more
complex samples with a variety of computing and networking capabilities.
On the other hand, malware authors have clearly benefited from newer
and richer integrated development environments (IDEs), frameworks, and libraries. This
explains the increasing modularity seen in the most recent samples--and,
especially, the rise of complex, multi-language malware projects that
would be otherwise unmanageable.

One interesting question 
is whether this trend will hold in time. If so, we could soon
see malware specimens with more than 1 million SLOC. 
To translate these numbers into real-world examples, in
the near future we could witness malware samples exceeding 
three times in size open source projects like Git or the Apache 
web server (see Table \ref{table:metricsSW}).
However, evolving into large pieces of software will surely
involve a higher amount of vulnerabilities and defects. This
has been already observed (and exploited), e.g., in~\cite{ekhunter}
and~\cite{decomposition}. 
In addition, such evolution requires larger efforts and thus possibly 
larger development teams. 
While we observe the trend we have not examined in 
detail those development teams. 
For this, we could apply authorship attribution techniques for 
source code~\cite{Frantzeskou2008examining,Caliskan2015deanonymzing}. 
More generally, the results shown in this paper provide
quantified evidence of how malware development
has been progressively transforming into a fully
fledged industry.

\section{Related Work}
\label{sec:related}

While malware typically propagates as binary code, 
some malware families have distributed themselves as source code. 
Arce and Levy performed an analysis of the 
Slapper worm source code~\cite{Arce2003analysis}, 
which upon compromising a host would upload 
its source code, compile it using gcc, and run the compiled executable. 
In 2005, Holz~\cite{Holz2005short} performed an analysis of the botnet 
landscape that describes how the source code availability of the 
Agobot and SDBot families lead to numerous variants of those families 
being created.

Barford and Yegneswaran~\cite{Barford2007inside} 
argue that we should develop a foundational understanding of
the mechanisms used by malware and that this can be achieved 
by analyzing malware source code available on the Internet. 
They analyze the source code of 4 IRC botnets 
(Agobot, SDBot, SpyBot, and GTBot) along 
7 dimensions: botnet control mechanisms, 
host control mechanisms, propagation, exploits, delivery mechanisms, 
obfuscation, and deception mechanisms.

Other works have explored the source code of exploit kits 
collected from underground forums and markets. 
Exploit kits are software packages installed on Web servers 
(called exploit servers) that try to compromise their visitors by 
exploiting vulnerabilities in Web browsers and their plugins.
Different from client malware, exploit kits are distributed as 
(possibly obfuscated) source code.
Kotov and Massacci~\cite{Kotov13anatomy} analyzed the source code of 
30 exploit kits collected from underground markets finding that 
they make use of a limited number of vulnerabilities.
They evaluated characteristics such as evasion, traffic statistics, and 
exploit management.
Allodi et al.~\cite{malwarelab} followed up on this research 
by building a malware lab to experiment with the exploit kits.
Eshete and Venkatakrishnan describe WebWinnow~\cite{webwinnow} a 
detector for URLs hosting an exploit kit, 
which uses features drawn from 40 exploit kits they installed in 
their lab.
Eshete et al. follow up this research line with EKHunter~\cite{ekhunter}
a tool that given an exploit kit finds vulnerabilities it may contain, 
and tries to automatically synthesize exploits for them. 
EKHunter finds 180 in 16 exploit kits (out of 30 surveyed), and 
synthesizes exploits for 6 of them.
Exploitation of malicious software was previously demonstrated 
by Caballero et al.~\cite{decomposition} directly on the binary code 
of malware samples installed in client machines.

The problem of detecting duplicated or cloned code was first approached
using simple text-matching solutions. The technique described in
\cite{baker1995finding} consists in a pairwise comparison among source
code files looking for a coincidence. While this allows to find exact copies
of source code, it does not scale well and may incur in performance issues.
In any case, note that text-matching approaches require a preliminary
normalization step such as the one used in this work.
A second group of techniques rely on data structures such as graphs or trees
to represent the syntactic structure of the 
programs~\cite{baxter1998clone,mayrand1996experiment}, together with an appropriate
similarity measure among them. Other works have proposed solutions based
on a lexical analysis of source files. These techniques convert the source
code sentences into lists of tokens, which are then compared to detect
duplicated subsequences~\cite{kamiya2002ccfinder,li2006cp}.

In the case of code sharing in malware, most existing work has
focused on binary objects~\cite{huang2017binsequence,jang2009bitshred}.
Even though the results reported are reasonable, one potential limitation
of such works is that modern compilers can perform different optimization
and cleaning tasks (e.g., loop unraveling, symbol stripping, etc.) 
to generate optimal binary objects in terms of size and memory consumption.
This could end up altering the structure of the original code and deleting
many valuable and meaningful artifacts \cite{rosenblum2010extracting}.
Contrarily, working directly with the original source code gives us more
precise insights on the functionality that is more frequently reused across
samples.

\section{Conclusion}
\label{sec:conclusion}
In this paper, we have presented a study on the evolution of malware source
code over the last four decades, 
as well as a study of source code reuse among malware families.
We have gathered and analyzed a dataset of 456 samples, which to our knowledge
is the largest of this kind studied in the literature. Our focus on software
metrics is an attempt to quantify properties both of the code itself and its
development process. The results discussed throughout the paper provide a
numerical evidence of the increase in complexity suffered by malicious code
in the last years and the unavoidable transformation into an engineering
discipline of the malware production process.

\section*{Acknowledgments}
This work was supported by the Spanish Government through 
MINECO grants SMOG-DEV (TIN2016-79095-C2-2-R) and 
DEDETIS (TIN2015-7013-R), and by the Regional Government of Madrid through 
grants CIBERDINE (S2013/ICE-3095) and N-GREENS (S2013/ICE-2731).

{
\bibliographystyle{IEEEtran}
\bibliography{refs}

}

\begin{IEEEbiography}
[{\includegraphics[width=1in,height=1.25in,clip,keepaspectratio]{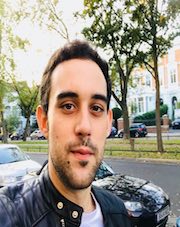}}]%
{Alejandro Calleja}
is a Ph.D. candidate in the Computer Security Lab (COSEC) at the Department of Computer Science and Engineering of Universidad Carlos III de Madrid under the supervision of Dr. Juan Tapiador. He holds a B.Sc and a M.Sc in computer science from Universidad Carlos III de Madrid. His main research line is automatic malware source code analysis. He is also also interested in several topics related with information and systems security, such as security in smartphone and mobile devices, security in embedded devices, computer forensics and reverse engineering.
\end{IEEEbiography}

\begin{IEEEbiography}[{\includegraphics[width=1in,height=1.25in,clip,keepaspectratio]{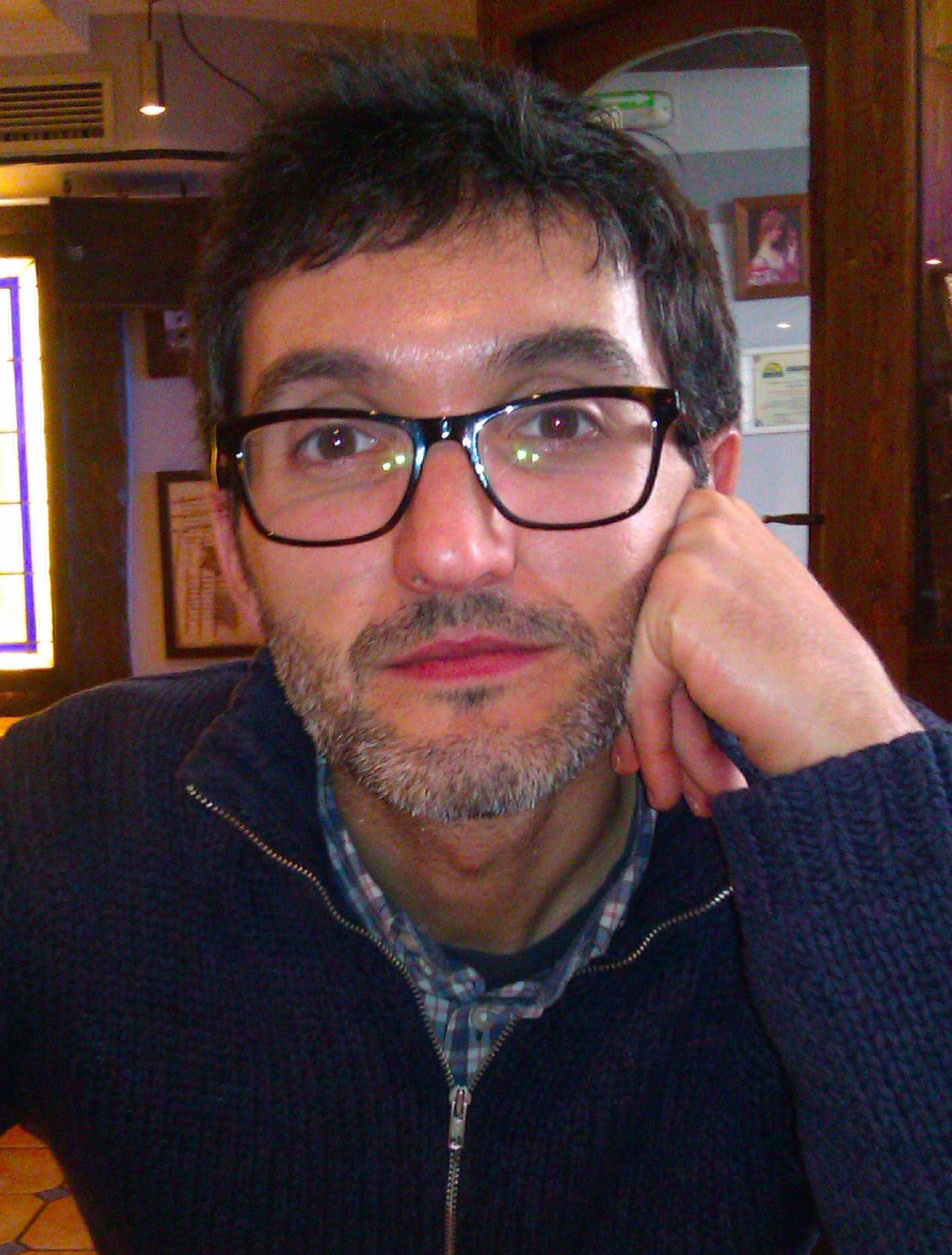}}]%
{Juan Tapiador}
Juan Tapiador is Associate Professor in the Department of Computer Science at Universidad Carlos III de Madrid, Spain, where he leads the Computer Security Lab. Prior to joining UC3M, I worked at the University of York, UK. His research focuses on various security issues of systems, software and networks, including malware analysis, attack modeling, anomaly and intrusion detection. He is Associate Editor for Computers \& Security (COSE) and has served in the technical committee of several venues in computer security, such as ACSAC, ACNS, DIMVA, ESORICS and AsiaCCS.
\end{IEEEbiography}

\begin{IEEEbiography}[{\includegraphics[width=1in,height=1.25in,clip,keepaspectratio]{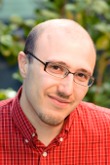}}]
{Juan Caballero}
Juan Caballero is Deputy Director and Associate Research Professor at the IMDEA Software Institute in Madrid, Spain. His research addresses security issues in systems, software, and networks. One of his focus is the analysis of malware and cyberattacks. He received his Ph.D. in Electrical and Computer Engineering from Carnegie Mellon University, USA. His research regularly appears at top security venues and has won two best paper awards at the USENIX Security Symposium, one distinguished paper award at IMC, and the DIMVA Most Influential Paper 2009-2013 award. He is an Associate Editor for ACM Transactions on Privacy and Security (TOPS). He has been program chair or co-chair for ACSAC, DIMVA, DFRWS, and ESSOS. He has been in the technical committee of the top venues in computer security including IEEE S\&P, ACM CCS, USENIX Security, NDSS, WWW, RAID, AsiaCCS, and DIMVA
\end{IEEEbiography}


\end{document}